\documentclass [12pt,eqsecnum,amsfonts,amssymb,aps]{revtex4}

\input epsf

\usepackage[usenames]{color}
\usepackage{graphicx}
\usepackage{bm}
\usepackage{rotating}

\newcommand{\beq}{\begin{equation}}
\newcommand{\eeq}{\end{equation}}
\newcommand{\beqs}{\begin{eqnarray}}
\newcommand{\eeqs}{\end{eqnarray}}

\begin{document}

\baselineskip 6.0mm

\title{Study of Exponential Growth Constants of Directed Heteropolygonal 
 Archimedean Lattices} 

\author{Shu-Chiuan Chang$^a$ and Robert Shrock$^b$}

\affiliation{(a) \ Department of Physics, National Cheng Kung University,
Tainan 70101, Taiwan} 

\affiliation{(b) \ C. N. Yang Institute for Theoretical Physics and
Department of Physics and Astronomy \\
Stony Brook University, Stony Brook, NY 11794, USA }

\bigskip
\bigskip

\begin{abstract}

  We infer upper and lower bounds on the exponential growth constants
  $\alpha(\Lambda)$, $\alpha_0(\Lambda)$, and $\beta(\Lambda)$ describing the
  large-$n$ behavior of, respectively, the number of acyclic orientations,
  acyclic orientations with a unique source vertex, and totally cyclic
  orientations of arrows on bonds of several $n$-vertex heteropolygonal
  Archimedean lattices $\Lambda$. These are, to our knowledge, the best bounds
  on these growth constants. The inferred upper and lower bounds on the growth
  constants are quite close to each other, which enables us to derive rather
  accurate values for the actual exponential growth constants. Combining our
  new results for heteropolygonal Archimedean lattices with our recent results
  for homopolygonal Archimedean lattices, we show that the exponential growth
  constants $\alpha(\Lambda)$, $\alpha_0(\Lambda)$, and $\beta(\Lambda)$ on
  these lattices are monotonically increasing functions of the lattice
  coordination number. Comparisons are made with the corresponding growth
  constants for spanning trees on these lattices. Our findings provide further
  support for the Merino-Welsh and Conde-Merino conjectures.

\end{abstract}

\maketitle

\newpage                                                                      

\pagestyle{plain}
\pagenumbering{arabic}


\section{Introduction}
\label{intro_section}

In this paper we continue our study of the exponential growth constants
$\alpha(\Lambda)$, $\alpha_0(\Lambda)$, and $\beta(\Lambda)$ describing the
large-$n$ behavior of, respectively, the number of acyclic orientations,
acyclic orientations with a unique source vertex, and totally cyclic
orientations of arrows on bonds of $n$-vertex Archimedean lattices $\Lambda$.
In Ref. \cite{ac} we inferred upper and lower bounds on these exponential
growth constants for several lattices, including the three homopolygonal
Archimedean lattices: square ($sq$), triangular ($tri$), and honeycomb ($hc$).
For each growth constant and lattice, our upper and lower bounds in \cite{ac}
were quite close to each other, which allowed us to obtain reasonably precise
values for the actual exponential growth constants. We also presented exact
values for $\alpha(tri)$, $\alpha_0(tri)$, and $\beta(hc)$. In the present
paper we extend this analysis to heteropolygonal Archimedean lattices.

We begin with some definitions and background. We refer the reader to our
previous paper \cite{ac} for more details and to \cite{graphtheory} for general
discussions of mathematical graph theory. A graph $G=(V,E)$ is defined by its
vertex and edge sets $V$ and $E$.  We denote $n(G)=|V|$, $e(G)=|E|$, $fc(G)$,
and $k(G)$ as the number of vertices (=sites), edges (= bonds), faces, and
connected components of $G$, respectively. We will focus on planar lattice
graphs with various boundary conditions in the longitudinal and transverse
directions, taken as the $x$ and $y$ directions, respectively.  The degree of a
vertex in a graph is the number of edges that connect to it.  A graph whose
vertices have the same degree $\Delta$ is termed a $\Delta$-regular graph.  An
Archimedean lattice is a uniform tiling of the plane with one or more types of
regular polygons, such that all vertices are equivalent \cite{gsbook}. A graph
that is a finite section of an Archimedean lattice with doubly periodic
boundary conditions is a $\Delta$-regular graph, and we will use this term also
in the limit $n(\Lambda) \to \infty$ (where it is synonymous with the lattice
coordination number), denoting it as $\Delta(\Lambda)$.  If an Archimedean
lattice is comprised of only a single type of regular polygon, it is termed
homopolygonal, while if it is comprised of two or more different types of
regular polygons, it is termed heteropolygonal. Owing to the equivalence of all
vertices of an Archimedean lattice $\Lambda$, it may be defined by the
ordered sequence of regular polygons that one traverses in a circuit
around any vertex:
\beq
\Lambda = (\prod_i p_i^{a_i}) \ , 
\label{arch}
\eeq
where the $i$'th polygon has $p_i$ sides and appears $a_i$ times contiguously
in the sequence (it can also occur non-contiguously). The total number of
occurrences of the polygon $p_i$ in the above sequence is denoted as
$a_{i,s}$. The number of polygons of type $p_i$ per vertex is $\nu_{p_i} =
a_{i,s}/p_i$. There are eleven Archimedean lattices, listed in Eqs.
(\ref{arch_degree3})-(\ref{arch_degree6}) in the Appendix. Of these, three are
homopolygonal, namely $(4^4)$ (square), $(3^6)$ (triangular), and $(6^3)$
(honeycomb), and the rest are heteropolygonal.

Given a graph $G$, we assign an arrow to each edge of $G$, thus defining a
directed graph (also called a digraph), $D(G)$. Since the arrow on each edge
has two possible orientations, there are $N_{eo}(G) = 2^{e(G)}$ possible
orientations of these arrows on edges of $G$ (where the subscript $eo$ stands
for ``edge orientations'').  An interesting and fundamental question in graph
theory concerns the numbers of certain subsets of the $N_{eo}(G)$ edge (arrow)
orientations and how these numbers grow as $n(G) \to \infty$.  Here we focus on
three classes of arrow orientations. We restrict our analysis of these subsets
of arrow orientations to connected graphs $G$, so $k(G)=1$; this does not
entail any loss of generality. A directed cycle on a directed graph $D(G)$ is
defined as a set of arrows on edges forming a cycle (circuit) such that, as one
traverses the cycle in a given direction, all of the arrows point in the
direction of motion. An acyclic orientation of the arrows on edges of $D(G)$ is
one in which there are no directed cycles. We denote $a(G)$ as the number of
acyclic orientations on the graph $G$. Now consider a given vertex in $G$. One
may enumerate the number of acyclic orientations on $G$ for which this, and
only this, vertex is a source vertex, i.e. has outgoing arrows on all edges
connected to it. This number is actually independent of the choice of the
vertex, and is denoted $a_0(G)$.  Among the various orientations of the arrows
on edges of $G$, some have the property that the arrow on each edge is a member
of a directed cycle. These are called totally cyclic (edge) orientations.  The
number of these, denoted as $b(G)$, constitutes a third basic quantity of
interest.  We restrict our analysis to graphs without loops (i.e., edges that
connect a vertex to itself), since if a graph $G$ has a loop, then $a(G)$ and
$a_0(G)$ both vanish identically.  In order to have a minimal measure of
totally cyclic orientations, we also restrict our analysis to graphs without
multiple edges, since one can increase $b(G)$ arbitrarily by replacing single
edges by multiple edges in a given graph.

The numbers $a(G)$ and $a_0(G)$ can be calculated from a knowledge of the
chromatic polynomial $P(G,q)$ of $G$, which enumerates the number of
assignments of $q$ colors to the vertices of $G$ satisfying the condition that
the colors on any two adjacent vertices (i.e., vertices connected by an edge)
are different \cite{chrompoly}. Such a color assignment is called a proper
$q$-coloring of (the vertices of) $G$. The minimum value of $q$ for which one
can perform a proper $q$-coloring of $G$ is the chromatic number of $G$,
denoted $\chi(G)$. The chromatic polynomial always has an
overall factor of $q$, so that one can define a reduced polynomial $P_r(G,q)
\equiv q^{-1} P(G,q)$. Although the chromatic polynomial $P(G,q)$ enumerates
proper $q$-colorings for positive integer values of $q$, it is also
well-defined for other values of $q$. Specifically, $a(G) = (-1)^{n(G)} \,
P(G,-1)$ \cite{stanley73} and $a_0(G) = (-1)^{n(G)-1} \, P_r(G,0)$
\cite{greene_zaslavsky83}.  As discussed
in the Appendix, the chromatic polynomial is a special case of an important
(two-variable) graph-theoretic function, namely the Tutte polynomial,
$T(G,x,y)$ \cite{tutte54}-\cite{welsh_merino2000}, \cite{graphtheory}, and
equivalent expressions for $a(G)$ and $a_0(G)$ are $a(G) = T(G,2,0)$
\cite{stanley73} and $a_0(G) = T(G,1,0)$
\cite{greene_zaslavsky83,gebhard_sagan99}.
Some previous studies of acyclic orientations on
square-lattice graphs include \cite{merino_welsh99}-\cite{ka3}. 
  The number $b(G)$ can be determined
in terms of the flow polynomial $F(G,q)$ evaluated at $q=-1$ or equivalently in
terms of the Tutte polynomial as $b(G) = T(G,0,2)$
\cite{lasvergnas77,lasvergnas80}.

For a wide class of families of lattice strip graphs $G$ of a given width and
arbitrarily great length, with certain longitudinal and transverse boundary
conditions, the quantities $a(G)$, $a_0$, and $b(G)$ grow exponentially rapidly
as a function of $n(G) \equiv n$ as $n \to \infty$. Let $\{ G \}$ denote the
formal limit of a given family of lattice strip graphs as $n \to \infty$.  One
thus defines exponential growth constants (EGCs) that describe the asymptotic
growth of these quantities as $n \to \infty$:
\beq
\alpha(\{ G \}) = \lim_{n \to   \infty} [a(G)]^{1/n} \ , 
\label{alpha}
\eeq
\beq
\alpha_0(\{ G \}) = \lim_{n \to \infty} [a_0(G)]^{1/n} \ , 
\label{alpha0}
\eeq
and
\beq
\beta(\{ G \}) = \lim_{n \to \infty} [b(G)]^{1/n} \ . 
\label{beta}
\eeq
A recursive family of graphs is a family such that the $(m+1)'th$ member of the
family can be obtained from the $m$'th member by adding a copy of a given
subgraph \cite{bds}.  An example is a family of lattice strips with a fixed
width $L_y$ and variable length $m$, together with some specified set of
longitudinal and transverse boundary conditions. For a recursive family of
graphs, in particular, a family of strips of some lattice $\Lambda$, the Tutte
polynomial $T(G,x,y)$, or equivalently, the Potts model partition function
$Z(G,q,v)$ (see Eqs. (\ref{z}), (\ref{t}) and (\ref{zt}) in the Appendix), and
hence also the chromatic polynomial, can be written as a sum of $m$'th powers
of certain functions, the set of which is generically denoted $\{ \lambda \}$,
that depend on the type of lattice $\Lambda$, the strip width $L_y$, and the
boundary conditions, but not on the length, $m$. In the infinite-length limit
$m \to \infty$, the $\lambda$ function with the largest magnitude dominates the
sum. Hence, when calculating the exponential growth constants in the limit of
infinitely long finite-width strips, it is only necessary to calculate the
dominant $\lambda$ function.  From our previous calculations of chromatic and
Tutte/Potts polynomials for a variety of lattice strip graphs, we know what
these dominant $\lambda$ functions are at the values $(q,v)=(-1,-1)$, $(0,-1)$,
and $(-1,1)$ (or equivalently, in terms of Tutte variables, $(x,y)=(2,0)$,
$(1,0)$, and $(0,2)$) needed to evaluate $\alpha( \{ G \})$, $\alpha_0( \{ G
\})$, and $\beta( \{ G \})$, respectively.

Because the Tutte polynomial $T(G,x,y)$ 
is equivalent to the partition function of the $q$-state Potts model in
statistical mechanics, $Z(G,q,v)$, with $x=1+(q/v)$ and $y=v+1$, 
the quantities $a(G)$, $a_0(G)$, and
$b(G)$ and the associated exponential growth constants $\alpha(\{ G \})$, 
$\alpha_0(\{ G \})$, and $\beta(\{ G \})$ have interesting connections 
with physical quantities, albeit at different values of $q$ than one 
studies in physical situations. Specifically, 
$\alpha(\{G\}) = |W(\{G\},-1)|$, 
$\alpha_0(\{G\}) = |W(\{G\},0)|$, and 
$\beta(\{G\}) = e^{|f(\{G\},-1,1)|}$, where $W(\{ G \},q)$ is the
zero-temperature degeneracy per vertex of the Potts antiferromagnet and 
$f(\{G \},q,v)$ is the dimensionless free energy per vertex of the Potts 
model, defined in Eqs. (\ref{w}) and (\ref{f}).  
It will also be of interest to compare these numbers with the exponential
growth constant $\tau(\{G\})$ describing the asymptotic growth of the number of
spanning trees on $G$ as $n(G) \to \infty$. 

For a $\Delta$-regular graph $G$, $N_{eo}(G)$, the number of edge orientations,
is given by the formula above with $e(G) = \Delta(G) \, n(G)/2$, i.e.,
$N_{eo}=2^{\Delta(G) n(G)/2}$.  We will also investigate some properties of
lattice strips of (heteropolygonal) Archimedean lattices and duals of
Archimedean lattices, some of which are not $\Delta$-regular graphs. For this
purpose, we define an effective vertex degree as $\Delta_{eff}(G)=2e(G)/n(G)$
and $\Delta_{eff}( \{ G \}) = \lim_{n \to \infty} 2e(G)/n(G)$, as in \cite{wn}.
This leads naturally to the definition of an exponential growth constant for
$N_{eo}$. For a $\Delta$-regular graph, this is $\epsilon(\{ G \}) \equiv
\lim_{n(G) \to \infty} [N_{eo}(G)]^{1/n(G)} = 2^{\Delta(\{ G \})/2}$ and for a
general graph, it is $\epsilon(\{ G \}) = 2^{\Delta_{eff}(\{ G \})/2}$.

This paper is organized as follows. In Section \ref{strip_section} we
illustrate the calculation of exponential growth constants with a specific
example of a strip of a heteropolygonal lattice.  In Section
\ref{methods_section} we briefly review our method of obtaining upper and lower
bounds on these exponential growth constants. Our resulting bounds are
presented in Section \ref{results_section}.  We discuss an
interesting connection with bounds on the zero-temperature degeneracy per
vertex of the Potts antiferromagnet in \ref{uw_section}. In Section
\ref{arch_dual_section} we present inferred upper bounds on exponential growth
constants for duals of Archimedean lattices.  Our results are given in Tables
\ref{lowerbounds_alpha_488_table}-\ref{uw_duals_table}.   A comparative
discussion is given in Section \ref{comparison_section}, and our conclusions
are stated in Section \ref{conclusion_section}.  Some useful results from graph
theory are included in an Appendix.


\section{Illustrative Calculations for Kagom\'e Strip} 
\label{strip_section}

Here we give a brief illustration of how exponential growth constants can be
calculated for strips of one type of heteropolygonal lattice, namely the $(3
\cdot 6 \cdot 3 \cdot 6)$ lattice. This is commonly called the kagom\'e
lattice, and we will use the abbreviation $kag$ for it. We take the
longitudinal and transverse directions to be the $x$ and $y$ directions,
respectively, and denote the boundary conditions as $(BC_y,BC_x)$.  (These
symbols $x$ and $y$ should not be confused with the variables $x$ and $y$ in
the Tutte polynomial $T(G,x,y)$ in Eq. (\ref{t}); the context will always make
clear the distinction.)  The boundary conditions $(BC_y,BC_x)$ are labelled as
(F,F) = free, (F,P) = cyclic, (P,F) = cylindrical, and (P,P) = toroidal.  Our
bounds for $\alpha( \{ G \})$, and $\alpha_0( \{ G \})$ are independent of
$BC_x$ but depend on $BC_y$, while the bounds for $\beta( \{ G \})$ depend on
both $BC_x$ and $BC_y$.

For definiteness, in this section we consider kagom\'e strips with (F,P), i.e.,
cyclic, boundary conditions.  The repeating subgraph in the first such strip of
this type consists of a hexagon and two adjacent triangles, say the upper and
lower left triangles (see Fig. 1(f) of \cite{strip}). This strip graph is
denoted $\{kagmin_m,cyc \}$, where the abbreviation $kagmin$ stands for ``
kagom\'e strip of minimal width'', and $cyc$ stands for cyclic.  A strip of
this type, with length $m$ of these subgraphs, has $n=5m$ vertices and $8m$
edges. It contains vertices with degree 3 and 4, and has an effective vertex
degree $\Delta_{eff} = 16/5 = 3.20$. The chromatic polynomial for this cyclic
strip was given in \cite{wcy,wcyc} in terms of a generating function. In an
equivalent form, we write it as
\beqs
P(kagmin_m,cyc,q) &=& (\lambda_{kag,0,+})^m + (\lambda_{kag,0,-})^m + 
(q-1)\Big [ (\lambda_{kag,1,+})^m + (\lambda_{kag,1,-})^m + 
(\lambda_{kag,1,3})^m \Big ] \cr\cr
&+& (q^2-3q+1)(\lambda_{kag,2})^m \ , 
\label{pkagmin}
\eeqs
where 
\beq
\lambda_{kag,0,\pm} = \frac{(q-2)}{2} \Big (T_{kag0} \pm \sqrt{R_{kag0}} \, 
\Big ) \ , 
\label{lamkag_d0_pm}
\eeq
\beq
T_{kag0}=q^4-6q^3+14q^2-16q+10 \ , 
\label{tkag0}
\eeq
\beq
R_{kag0} = q^8-12q^7+64q^6-200q^5+404q^4-548q^3+500q^2-292q+92 \ , 
\label{rkag0}
\eeq
\beq
\lambda_{kag,1,\pm} = \frac{1}{2}\Big (T_{kag1} \pm \sqrt{R_{kag1}} \, \Big ) 
\ , 
\label{lamkag_d1_pm}
\eeq
\beq
T_{kag1} = q^3-7q^2+19q-20 \ , 
\label{tkag1}
\eeq
\beq
R_{kag1} = q^6-14q^5+83q^4-278q^3+569q^2-680q+368 \ , 
\label{rkag1}
\eeq
\beq
\lambda_{kag,1,3} = (q-1)(q-2)^2 \ , 
\label{lamkag_d1_3}
\eeq
and
\beq
\lambda_{kag,2} = q-4 \ . 
\label{lamkag_d2}
\eeq
Evaluating $P(kagmin_m,cyc,q)$ at $q=-1$, we obtain 
\beqs
a(kagmin_m,cyc) &=& \Bigg [ \frac{3(47+\sqrt{2113})}{2} \Bigg ]^m + 
                  \Bigg [ \frac{3(47-\sqrt{2113})}{2} \Bigg ]^m \cr\cr
 &-&2 \Bigg [ \Bigg ( \frac{47+\sqrt{1993}}{2} \Bigg )^m + 
              \Bigg ( \frac{47-\sqrt{1993}}{2} \Bigg )^m + (18)^m \Bigg ]
+ 5^{m+1} \ . \cr\cr
&& 
\label{akagmin}
\eeqs
Taking the $m \to \infty$ limit yields the result 
\beqs
\alpha(\{ kagmin_{2 \times \infty} \}) = 
\Bigg [ \frac{3(47+\sqrt{2113} )}{2} \Bigg ]^{1/5} = 2.684630 \ .
\label{alpha_kagmin}
\eeqs
Similarly, calculating $a_0(mkag_m,cyc)$ and taking $m \to \infty$, we get
\beqs
\alpha_0(\{ kagmin_{2 \times \infty} \}) = [2(5+\sqrt{23} \, )]^{1/5} = 1.813069 \ .
\label{alpha0_kagmin}
\eeqs

A second type of cyclic strip graph of the kagom\'e lattice can be constructed
from the first by adjoining triangles to each of the edges of hexagons on one
side of the strip, say the upper side, so that the basic subgraph that repeats
$m$ times is a hexagon with three adjacent triangles, on the upper and lower
left of a given hexagon and above it. We denote this strip graph as $\{kag_m,cyc\}$.
It has $n=6m$ vertices and $10m$ edges.  It contains vertices with degrees 2,
3, and 4, and has an effective $\Delta_{eff} = 10/3 = 3.33$.  Knowing
$P(kagmin_m,cyc,q)$, an elementary calculation yields 
\beq
P(kag_m,cyc,q) = (q-2)^m P(kagmin_m,cyc,q) \ . 
\label{pkag}
\eeq
Hence, 
\beq
\alpha(kag_{2 \times \infty}) = 
3^{1/3} \, \Bigg ( \frac{47+\sqrt{2113}}{2} \Bigg )^{1/6} = 2.734789
\label{alpha_kag}
\eeq
and 
\beq
\alpha_0(kag_{2 \times \infty}) = 2^{1/3}(5+\sqrt{23} \, )^{1/6} = 1.842964 \ .
\label{alpha0_kag}
\eeq
The values in Eqs. (\ref{alpha_kag}) and (\ref{alpha0_kag}) are listed in Table
\ref{lowerbounds_alpha_kag_table}.  More generally, the results for
infinite-length finite-width strips of the kagom\'e lattice presented in Table
\ref{lowerbounds_alpha_kag_table} use a strip comprised of $L_y-1$ layers of
$kagmin$ glued to one layer of $kag$, i.e., they have triangles protruding on
one side, say the upper one, but a ``flat'' lower side.

A third type of cyclic strip graph of the kagom\'e lattice can be constructed
from the first by adjoining triangles to each of the edges of hexagons on both
the upper and the lower sides, so that the basic subgraph that repeats $m$
times is a hexagon with four adjacent triangles, on the upper and lower left of
a given hexagon and above and below it. We denote this strip graph as $kagt_m$,
where the $t$ in $kagt$ refers to the additional triangle subgraphs. It has $n=7m$ vertices, 
$12m$ edges and an effective vertex degree $\Delta_{eff}=24/7 = 3.429$.  
Another elementary calculation yields 
\beq
P(kagt_m,cyc,q) = (q-2)^m P(kag_m,cyc,q) = (q-2)^{2m}P(kagmin_m,cyc,q) \ . 
\label{pkagt}
\eeq
Hence, 
\beq
\alpha(\{ kagt \}_{2 \times \infty}) = 
3^{3/7} \, \Bigg ( \frac{47+\sqrt{2113}}{2} \Bigg )^{1/7} = 2.771190 
\label{alpha_kagt}
\eeq
and
\beq
\alpha_0(\{ kagt \}_{2 \times \infty}) = 2^{3/7}(5+\sqrt{23} \, )^{1/7} =
1.864619 \ . 
\label{alpha0_kagt}
\eeq

The fact that these exponential growth constants increase as the width of the
strip increases is consistent with the inference that these provide lower
bounds on $\alpha(kag)$ and $\alpha_0(kag)$.  This is the same type of behavior
that we showed for homopolygonal lattice strips in \cite{ka3,ac}.  Similar
illustrative calculations can be given for other strips.


\section{Methods for Calculation of Upper and Lower Bounds on Exponential
  Growth Constants} 
\label{methods_section}

In our previous studies \cite{ka3,ac} we showed that the resultant values of
$\xi(\Lambda,(L_y)_F \times \infty)$ and $\xi(\Lambda,(L_y)_P \times \infty)$
were monotonically increasing functions of $L_y$ for all of the widths $L_y$ of
homopolygonal lattices considered, where $\xi$ denotes any of the exponential
growth constants $\alpha$, $\alpha_0$, and $\beta$. Our results for
heteropolygonal lattice strips exhibit the same monotonicity, providing further
support for the inference that these quantities are lower bounds on the values
of the respective exponential growth constants for the infinite lattices. Our
results also provide further support for our earlier inference in \cite{ka3,ac}
that as $L_y \to \infty$, the values of $\xi(\Lambda,(L_y)_F \times \infty)$
and $\xi(\Lambda,(L_y)_P \times \infty)$ converge to the same unique value,
denoted $\xi(\Lambda)$, which is independent of the longitudinal and transverse
boundary conditions and thus characterizes the infinite lattice $\Lambda$.
Since the strips with periodic transverse boundary conditions (cylindrical or
toroidal) have no transverse boundary, the resulting values of the exponential
growth constants on finite-width, infinite-length strips should approach the
respective values for the infinite two-dimensional lattices more rapidly, and
we do observe this for strips of heteropolygonal lattices, as we did earlier in
\cite{ka3,ac} for strips of homopolygonal lattices.

As in \cite{ac}, as a quantitative measure the convergence of values of 
$\alpha(\Lambda,(L_y)_{BC_y} \times \infty)$ for consecutive values of strip
width to a constant limiting value, we define the ratio
\beq
R_{\alpha,\Lambda,(L_y+1)/L_y,BC_y} \equiv 
\frac{\alpha(\Lambda,(L_y+1)_{BC_y} \times \infty)}
     {\alpha(\Lambda,(L_y)_{BC_y} \times \infty)} \ . 
\label{ralpha}
\eeq
Just as was the case in \cite{ac} for homopolygonal lattice strips, we find
that this ratio approaches close to 1 even for modest values of the strip
widths. Our results for $\beta$ values in \cite{ac} 

In \cite{ac} we showed that ratios of the $\lambda$ functions for successive
strip widths provide an upper bound on the respective exponential growth
constants.  We refer the reader to \cite{ac} for this discussion. 
We proceed to present our results for the heteropolygonal Archimedean lattices
that we study. 


\section{Upper and Lower Bounds on Exponential Growth Constants on
  Heteropolygonal Archimedean Lattices} 
\label{results_section}

In this section we present upper and lower bounds that we have inferred for
$\alpha(\Lambda)$, $\alpha_0(\Lambda)$, and $\beta(\Lambda)$ on the
heteropolygonal Archimedean lattices $\Lambda$ that we study.  We obtain these
bounds via the calculations of infinite-length, finite-width strips of these
lattices, using methods discussed in \cite{ac} and reviewed in Section
\ref{methods_section}.


\subsection{$(4 \cdot 8^2)$ Lattice} 

We present our results for the $(4 \cdot 8^2)$ lattice in Tables
\ref{lowerbounds_alpha_488_table}-\ref{upperbounds_beta_488_table}.  In these
and later tables, because we utilize the entries with the highest values of the
width $L_y$ of strips for our upper and lower bounds, we list these to slightly
higher precision than the entries for smaller widths.  As is evident from these
tables and the others to be given below, we achieve very good precision in our
upper and lower bounds with modest values of $L_y$ for the lattice strips.
This was also true of our calculations on the homopolygonal Archimedean
lattices in \cite{ac}. From these new results, we infer the following upper and
lower bounds:
\beq
2.729704176 < \alpha((4 \cdot 8^2)) < 
2.730093140
\label{alpha_bounds_488}
\eeq
\beq
2.032649948 < \alpha_0((4 \cdot 8^2)) < 
2.077301063
\label{alpha0_bounds_488}
\eeq
and
\beq
2.080338691 < \beta((4 \cdot 8^2)) < 
2.107715225 \ . 
\label{beta_bounds_488}
\eeq
As was the case with our upper and lower bounds for the homopolygonal
Archimedean lattices in \cite{ac}, these bounds are quite close to each other,
which enables us to infer approximate values of the exponential growth
constants themselves. As a measure of this, for a general Archimedean lattice
$\Lambda$, we define the fractional difference
\beq
\frac{\xi_u(\Lambda)-\xi_\ell(\Lambda)}{\xi_{ave}(\Lambda)} \ ,  
\label{xi_fracdif}
\eeq
where $\xi(\Lambda)$ is any of the growth constants, $\alpha(\Lambda)$,
$\alpha_0(\Lambda)$, or $\beta(\Lambda)$ and $\xi_u(\Lambda)$, and
$\xi_\ell(\Lambda)$ are the corresponding upper ($u$) and lower ($\ell$) 
bounds. We further define the average value 
\beq
\xi_{ave}(\Lambda) = \frac{\xi_\ell(\Lambda)+\xi_u(\Lambda)}{2}  \ . 
\label{xi_average}
\eeq
For the $(4 \cdot 8^2)$ lattice, we have 
\beq
\frac{\alpha_u((4 \cdot 8^2))-\alpha_\ell((4 \cdot 8^2))}
     {\alpha_{ave}((4 \cdot 8^2))} = 1.425 \times 10^{-4}
\label{alpha_488_fracdif}
\eeq
\beq
\frac{\alpha_{0,u}((4 \cdot 8^2))-\alpha_{0,\ell}((4 \cdot 8^2))}
     {\alpha_{0,ave}((4 \cdot 8^2))} = 2.17 \times 10^{-2}
\label{alpha0_488_fracdif}
\eeq
and
\beq
\frac{\beta_u((4 \cdot 8^2))-\beta_\ell((4 \cdot 8^2))}
     {\beta_{ave}((4 \cdot 8^2))} = 1.31 \times 10^{-2} \ . 
\label{beta_488_fracdif}
\eeq

The interval separating the average value of $\xi(\Lambda)$ from the upper 
and lower bounds is 
\beq
\delta_{\xi(\Lambda)} \equiv \xi_u(\Lambda) - \xi_{ave}(\Lambda) = 
\xi_{ave}(\Lambda) - \xi_\ell(\Lambda) \ . 
\label{delta_xi}
\eeq
The approximate ($ap$) values of the exponential growth
constants, denoted $\xi_{ap}(\Lambda)$ for $\xi=\alpha, \ \alpha_0$, and
$\beta$, are given by 
\beq
\xi_{ap}(\Lambda) = \xi_{ave}(\Lambda) \pm \delta_{\xi(\Lambda)} \ . 
\label{xi_value}
\eeq
We calculate 
\beq
\alpha_{ap}((4 \cdot 8^2)) = 2.72990 \pm 
                              0.00019 
\label{alpha_est_488}
\eeq
\beq
\alpha_{0,ap}((4 \cdot 8^2)) = 2.055 \pm 
                                0.022
\label{alpha0_est_488}
\eeq
and
\beq
\beta_{ap}((4 \cdot 8^2)) =  2.094 \pm 
                              0.014 \ . 
\label{beta_est_488}
\eeq
%


\subsection{$(3 \cdot 6 \cdot 3 \cdot 6)$ (Kagom\'e) Lattice }

We present our results for the $(3 \cdot 6 \cdot 3 \cdot 6)$ (kagom\'e) lattice
in Tables \ref{lowerbounds_alpha_kag_table}-\ref{upperbounds_beta_kag_table}. 
From these we infer the following upper and lower bounds:
\beq
3.249059070 < \alpha(kag) < 
3.265737199
\label{alpha_bounds_kag}
\eeq
\beq
2.481974714 < \alpha_0(kag) < 
2.632503652
\label{alpha0_bounds_kag}
\eeq
and
\beq
3.415032724 < \beta(kag) < 
3.549454037 \ . 
\label{beta_bounds_kag}
\eeq

The fractional differences $[\xi_u(kag)-\xi_\ell(kag)]/\xi_{ave}(kag)$ 
are of order $10^{-2}$ for $\xi=\alpha, \ \alpha_0, \ \beta$. We compute 
\beq
\alpha_{ap}(kag) = 3.2574 \pm 
                    0.0083
\label{alpha_est_kag}
\eeq
\beq
\alpha_{0,ap}(kag) = 2.557 \pm 
                      0.075
\label{alpha0_est_kag}
\eeq
and
\beq
\beta_{ap}(kag) = 3.482 \pm 
                   0.067 \ . 
\label{beta_est_kag}
\eeq
%


\subsection{$(3^3 \cdot 4^2)$ Lattice }

We present our results for the $(3^3 \cdot 4^2)$ lattice in Tables
\ref{lowerbounds_alpha_33344_table}- \ref{upperbounds_beta_33344_table}. For
this lattice there are two different ways to choose the longitudinal direction
for the strips. Referring to Fig. 1(a) in \cite{sti}, we can choose either
$L_y$ in the vertical direction and $L_x$ in the horizontal direction, or vice
versa. We give results for both cases and use the most stringent ones (the
largest lower bound and the smallest upper bound) for our results. We obtain
\beq
3.922582062 < \alpha((3^3 \cdot 4^2)) < 
3.956121920
\label{alpha_bounds_33344}
\eeq
\beq
3.142411228 < \alpha_0((3^3 \cdot 4^2)) < 
3.298937504
\label{alpha0_bounds_33344}
\eeq
and
\beq
5.262880165 < \beta((3^3 \cdot 4^2)) < 
5.362606470 \ . 
\label{beta_bounds_33344}
\eeq
The fractional differences
$[\xi_u(\Lambda)-\xi_\ell(\Lambda)]/\xi_{ave}(\Lambda)$, where
$\xi= \alpha, \ \alpha_0, \ \beta$,  are of order 
$10^{-2}$ for this $\Lambda=(3^3 \cdot 4^2)$ lattice. We find 
\beq
\alpha_{ap}((3^3 \cdot 4^2)) = 3.939 \pm 
                               0.017
\label{alpha_est_33344}
\eeq
\beq
\alpha_{0,ap}((3^3 \cdot 4^2)) = 3.221 \pm 
                                 0.078
\label{alpha0_est_33344}
\eeq
and
\beq
\beta_{ap}((3^3 \cdot 4^2)) =  5.313 \pm 
                               0.050 \ . 
\label{beta_est_33344}
\eeq
%


\subsection{$(3^2 \cdot 4 \cdot 3 \cdot 4)$ Lattice }
\label{33434_section}

We present our results for the $(3^2 \cdot 4 \cdot 3 \cdot 4)$ lattice
in Tables \ref{lowerbounds_alpha_33434_table}-
\ref{upperbounds_beta_33434_table}. We have 
\beq
3.922582062 < \alpha((3^2 \cdot 4 \cdot 3 \cdot 4)) < 
3.9563647405
\label{alpha_bounds_33434}
\eeq
\beq
3.142411228 < \alpha_0((3^2 \cdot 4 \cdot 3 \cdot 4)) < 
3.299213098
\label{alpha0_bounds_33434}
\eeq
and
\beq
5.264056522 < \beta((3^2 \cdot 4 \cdot 3 \cdot 4)) < 
5.360035653 \ . 
\label{beta_bounds_33434}
\eeq
We compute 
\beq
\alpha_{ap}((3^2 \cdot 4 \cdot 3 \cdot 4)) = 3.939 \pm 
                                             0.017
\label{alpha_est_33434}
\eeq
\beq
\alpha_{0,ap}((3^2 \cdot 4 \cdot 3 \cdot 4)) = 3.221 \pm 
                                               0.078
\label{alpha0_est_33434}
\eeq
and
\beq
\beta_{ap}((3^2 \cdot 4 \cdot 3 \cdot 4)) = 5.312 \pm 
                                            0.048 \ . 
\label{beta_est_33434}
\eeq
Evidently, the upper and lower bounds on $\xi((3^3 \cdot 4^2))$ are very close
or equal (to the indicated number of significant figures) to the corresponding
$\xi((3^2 \cdot 4 \cdot 3 \cdot 4))$, where here $\xi(\Lambda)$ denotes
$\alpha(\Lambda)$, $\alpha_0(\Lambda)$, or $\beta(\Lambda)$, and so are the
resultant approximate values. This presumably reflects the fact that a circuit
around any vertex on these lattices contains the same number of triangles
(namely three) and squares (namely two), and the only difference is the order
in which these appear in the traversal.


\subsection{Summary of Bounds from Strip Calculations}

In Table \ref{egc_values_table} we list the approximate values of
$\alpha_{ap}(\Lambda)$, $\alpha_{0,ap}(\Lambda)$, and $\beta_{ap}(\Lambda)$ for
these heteropolygonal Archimedean lattices, together with the corresponding
quantities for homopolygonal Archimedean lattices. For the cases where we
presented exact results in Ref. \cite{ac}, namely $\alpha(tri)$,
$\alpha_0(tri)$, and $\beta(hc)$, we list these instead of the approximate
values.  As we found in \cite{ac} for homopolygonal Archimedean lattices, so
also here we find, for these heteropolygonal Archimedean lattices, that our
bounds and the resultant values of $\alpha_{ap}(\Lambda)$,
$\alpha_{0,ap}(\Lambda)$, and $\beta_{ap}(\Lambda)$ are monotonically
increasing functions of vertex degree (i.e., lattice coordination number),
$\Delta(\Lambda)$. 

Applying similar techniques, we have also obtained bounds on exponential growth
constants for spanning forests and connected spanning subgraphs on
heteropolygonal Archimedean lattices. Recall that a spanning forest in a graph
$G$ is a spanning subgraph of $G$ that does not contain any circuits. Let us
denote the number of spanning forests of a graph $G$ as $N_{SF}(G)$ and define
$\phi(\{G\}) \equiv \lim_{n(G) \to \infty} [N_{SF}(G)]^{1/n(G)}$. Recall that
$N_{SF}(G)=T(G,2,1)$. As an illustration, for the $(4 \cdot 8^2)$ lattice, we
infer the bounds $2.779135 \le \phi((4 \cdot 8^2)) \le 2.779486$. Since the
upper and lower bounds are very close to each other, we determine the
approximate value to be $\phi((4 \cdot 8^2)) = 2.77931 \pm 0.00018$.  We will
present these results elsewhere.


\section{Connection with $W(\Lambda,q)$ Bounds for Archimedean Lattices 
$\Lambda$}
\label{uw_section}

The property that we find in our calculations, that $\alpha(\Lambda,L_y,free)$
and $\alpha_0(\Lambda,L_y,free)$ are monotonically increasing functions of
strip width for a strip of the lattice $\Lambda$, is the reverse of the
dependence on strip width that was found with $W(\Lambda,L_y,free,q)$ when the
latter function is evaluated at $q \ge \chi(\Lambda)$, i.e., the range of $q$
required for a proper $q$-coloring of the lattice $\Lambda$ \cite{bcc99}. We
interpret this as a consequence of the fact that $\alpha(\Lambda,L_y,free)$ and
$\alpha_0(\Lambda,L_y,free)$ involve the evaluation of $|W(\Lambda,q)|$ at
different values of $q$, namely $q=-1$ and $q=0$, respectively.

In \cite{ac} we noted our observation concerning analytic expressions that were
proved to be lower bounds on $W(\Lambda,q)$ for Archimedean lattices in
\cite{wn} (see also \cite{w}-\cite{ilb}) with $q \ge \chi(\Lambda)$, using a
coloring-matrix method that had been applied to derive a lower bound on
$W(sq,q)$ in \cite{biggscoloring}. The observation was that, for a given
Archimedean lattice $\Lambda$, if one sets $q=-1$ or $q=0$ in the analytic
expressions that had been proved in \cite{wn} to be lower bounds on
$W(\Lambda,q)$ for $q \ge \chi(\Lambda)$, then the resultant values are
consistent with being upper bounds on $\alpha(\Lambda)$ and
$\alpha_0(\Lambda)$, respectively. Therefore, we conjectured in 
\cite{ac} that these evaluations are, indeed, upper bounds on $\alpha(\Lambda)$
and $\alpha_0(\Lambda)$.  We recall that the lower bound on $W(\Lambda,q)$ that
was proved in \cite{wn}, where $\Lambda$ is an Archimedean lattice, is
\beq
W(\Lambda,q) \ge W(\Lambda,q)_\ell \ , 
\label{wwlow}
\eeq
where (see Eq. (4.11) of \cite{wn}) 
\beq
W \bigg ( (\prod_i p_i^{a_i} ),q \bigg )_\ell 
= \frac{\prod_i [D_{p_i}(q)]^{\nu_{p_i}}} {q-1} \ , 
\label{wlowform}
\eeq
where $\nu_{p_i}$ was defined below Eq. (\ref{arch}), 
\beq
D_n(q) = \sum_{s=0}^{n-2} (-1)^s {n-1 \choose s} q^{n-2-s} \ ,
\label{dn}
\eeq
and $q \ge \chi(\Lambda)$. 
The conjectured upper bounds on $\alpha(\Lambda)$ and $\alpha_0(\Lambda)$ 
are then as follows, where $\Lambda$ is an Archimedean lattice: 
\beq
\alpha(\Lambda) < \alpha_{u,w}(\Lambda) 
\label{alpha_lt_alphaw}
\eeq
and
\beq
\alpha_0(\Lambda) < \alpha_{0,u,w}(\Lambda) \ , 
\label{alpha0_lt_alpha0w}
\eeq
where
\beq
\alpha_{u,w}(\Lambda) = \frac{\prod_i |D_{p_i}(-1)|^{\nu_{p_i}}}{2} 
\label{alfup_uw}
\eeq
and
\beq
\alpha_{0,u,w}(\Lambda) = \prod_i |D_{p_i}(0)|^{\nu_{p_i}} \ . 
\label{alf0up_uw}
\eeq

We list these below for each of the eleven Archimedean lattices, in order of
increasing vertex degree, $\Delta(\Lambda)$, and for a given $\Delta(\Lambda)$,
in order of increasing girth, $g(\Lambda)$. To indicate their connection with
bounds on the $W$ function, we append a subscript $w$: 
\beq
\alpha_{u,w}((3 \cdot 12^2))= \frac{3^{1/3} \times (2047)^{1/6}}{2} = 
2.569587414
\label{wupperbound_alpha_31212}
\eeq
\beq
\alpha_{u,w}((4 \cdot 8^2)) = \frac{7^{1/4} \times (127)^{1/4}}{2} = 
2.730206175
\label{wupperbound_alpha_488}
\eeq
\beq
\alpha_{u,w}((4 \cdot 6 \cdot 12)) = \frac{7^{1/4} \times (31)^{1/6} \times 
(2047)^{1/12}}{2} = 
2.721017178
\label{wupperbound_alpha_4612}
\eeq
\beq
\alpha_{u,w}((6^3)) \equiv \alpha(hc) = \frac{\sqrt{31}}{2} = 
2.783882181
\label{wupperbound_alpha_hc}
\eeq
\beq
\alpha_{u,w}((3 \cdot 6 \cdot 3 \cdot 6)) \equiv \alpha_{u,w}(kag) = 
\frac{3^{2/3} \times (31)^{1/3}}{2}= 
3.2671675385
\label{wupperbound_alpha_kag}
\eeq
\beq
\alpha_{u,w}((3 \cdot 4 \cdot 6 \cdot 4)) =
\frac{3^{1/3} \times 7^{1/2} \times (31)^{1/6}}{2} = 
3.381580457 
\label{wupperbound_alpha_3464}
\eeq
\beq
\alpha_{u,w}((4^4)) \equiv \alpha_{u,w}(sq) = \frac{7}{2} 
\label{wupperbound_alpha_sq}
\eeq
\beq
\alpha_{u,w}((3^3 \cdot 4^2))= 
\alpha_{u,w}((3^2 \cdot 4 \cdot 3 \cdot 4))=\frac{3\sqrt{7}}{2} = 
3.968626967
\label{wupperbound_alpha_33344}
\eeq
\beq
\alpha_{u,w}((3^4 \cdot 6))= \frac{3^{4/3} \times (31)^{1/6}}{2} =
3.834351826
\label{wupperbound_alpha_33336}
\eeq
\beq
\alpha_{u,w}((3^6)) \equiv \alpha_{u,w}(tri) = \frac{9}{2} 
\label{wupperbound_alpha_tri}
\eeq
and
\beq
\alpha_{0,u,w}((3 \cdot 12^2))= 2^{1/3} \times (11)^{1/6} = 
1.878922121 
\label{wupperbound_alpha0_31212}
\eeq
\beq
\alpha_{0,u,w}((4 \cdot 8^2))= 3^{1/4} \times 7^{1/4} = 
2.140695143
\label{wupperbound_alpha0_488}
\eeq
\beq
\alpha_{0,u,w}((4 \cdot 6 \cdot 12))=
3^{1/4} \times 5^{1/6} \times (11)^{1/12}= 
2.101638609
\label{wupperbound_alpha0_4612}
\eeq
\beq
\alpha_{0,u,w}((6^3)) \equiv \alpha_{0,u,w}(hc) = \sqrt{5} = 
2.2360679775
\label{wupperbound_alpha0_hc}
\eeq
\beq
\alpha_{0,u,w}((3 \cdot 6 \cdot 3 \cdot 6)) \equiv 
\alpha_{0,u,w}(kag) = 2^{2/3} \times 5^{1/3} = 
2.714417617
\label{wupperbound_alpha0_kag}
\eeq
\beq
\alpha_{0,u,w}((3 \cdot 4 \cdot 6 \cdot 4))=
2^{1/3} \times 3^{1/2} \times 5^{1/6} = 
2.853638528 
\label{wupperbound_alpha0_3464}
\eeq
\beq
\alpha_{0,u,w}((4^4)) \equiv \alpha_{0,u,w}(sq) = 3
\label{wupperbound_alpha0_sq}
\eeq
\beq
\alpha_{0,u,w}((3^3 \cdot 4^2)) = 
\alpha_{0,u,w}((3^2 \cdot 4 \cdot 3 \cdot 4) = 2\sqrt{3} =
3.464101615
\label{wupperbound_alpha0_33344}
\eeq
\beq
\alpha_{0,u,w}((3^4 \cdot 6))= 2^{4/3} \times 5^{1/6} = 
3.295097945
\label{wupperbound_alpha0_33336}
\eeq
and 
\beq
\alpha_{0,u,w}((3^6)) \equiv \alpha_{u,w,0}(tri) = 4 \ . 
\label{wupperbound_alpha0_tri}
\eeq

Using Eq. (4.18) of Ref. \cite{wn}, we can slightly improve the suggested upper
bounds for the $(4 \cdot 6 \cdot 12)$ lattice as follows: 
\beq
\alpha((4 \cdot 6 \cdot 12))_{u,w} = \frac{(42170569)^{1/12}}{2^{2/3}} =
2.721014094
\label{upperbound2_alpha_4612}
\eeq
and
\beq
\alpha_0((4 \cdot 6 \cdot 12))_{u,w}=3^{1/6} \times (805)^{1/12} = 
2.097344955 \ . 
\label{upperbound2_alpha0_4612}
\eeq
These numerical values are listed in Table \ref{uw_table}.  In this table we
also list the exact values of $\tau(\Lambda)$ from \cite{sti},
\cite{wu77}-\cite{std} (to the indicated number of significant figures). 


\section{Connection with $W(\Lambda_{dual},q)$ Bounds on Duals of Archimedean 
Lattices}
\label{arch_dual_section} 

\subsection{General} 

To each Archimedean lattice $\Lambda=(\prod_i p_i^{a_i})$, there corresponds a
planar dual lattice $\Lambda_{dual}$ obtained by mapping the vertices and faces
of $\Lambda$ to the faces and vertices, respectively, of $\Lambda_{dual}$. Just
as all of the vertices of an Archimedean lattice are equivalent, all of the
faces of the dual of an Archimedean lattice, i.e., the polygons of which it is
comprised, are equivalent. The dual Archimedean lattice $\Lambda_{dual}$
is defined by the ordered product of degrees of vertices that one traverses in
a circuit around the boundary of any face, 
\beq
\Lambda_{dual} = \Big [ \prod_i \Delta(v_i)^{b_i} \Big ] \ , 
\label{laves}
\eeq
where in the above product, the notation $\Delta(v_i)^{b_i}$ indicates that the
vertex $v_i$ with degree $\Delta(v_i)$ occurs contiguously $b_i$ times (and can
also occur noncontiguously) in the circuit. We define $b_{i,s} = \sum_i b_i$.
The polygons of which $\Lambda_{dual}$ is comprised have $p=b_{i,s}$ sides.
For notational clarity, square brackets are used in Eq. (\ref{laves}) for 
dual Archimedean lattices, while parentheses are used in Eq. (\ref{arch}) for
Archimedean lattices, so, e.g., the Archimedean lattice $(3 \cdot 12^2)$ has,
as its (planar) dual, the $[3 \cdot 12^2]$ lattice, and so forth for
others. The dual of an Archimedean lattice is also Archimedean if and only if
the original lattice is homopolygonal. Specifically, the duality transformation
maps the square lattice to an isomorphic copy of itself, and interchanges the
triangular and honeycomb lattices, i.e., $[4^4]=(4^4) = (sq)$, $[3^6]=(6^3) =
(hc)$, and $[6^3]=(3^6) = (tri)$.  The duals of heteropolygonal Archimedean
lattices are not, themselves, Archimedean, since they are heterovertitial,
i.e., contain vertices of different degrees.  One can still define an effective
vertex degree for these heterovertitial Archimedean dual lattices. 
Using $\Delta_{eff}(\{G\}) = \lim_{n(G) \to \infty} 2e(G)/n(G)$ as discussed in
the introduction, one has \cite{wn}
\beq
\Delta_{eff}(\Lambda_{dual}) = \nu_p \,  p = \frac{2p}{p-2} \ , 
\label{deltaeff_lambdadual}
\eeq
where 
\beq
\nu_p=\Bigg [\sum_i \frac{b_{i,s}}{\Delta(v_i)}\Bigg ]^{-1}=\frac{2}{p-2} \ . 
\label{nup}
\eeq
Note that for a lattice that is $\Delta$-regular, the effective vertex degree
just reduces to the uniform vertex degree.  In particular, for the duals of 
homopolygonal ($hp$) Archimedean lattices (which are thus $\Delta$-regular), 
$\Delta_{eff}(\Lambda_{dual,hp})=\Delta(\Lambda_{dual,hp})$.


\subsection{Values of $\beta_{u,w}(\Lambda_{dual})$} 

Using our inferred upper bounds on $\alpha(\Lambda)$ in conjunction with the
duality relation (\ref{beta_dual}) and the values of $\nu(\Lambda)$ given below
in Eqs. (\ref{nu_hetero_deg3})-(\ref{nu_hetero_deg5}), we thus obtain inferred
upper bounds on $\beta(\Lambda_{dual})$ for Archimedean dual lattices. Since
the conjectured upper bounds on $\alpha(\Lambda)$ and $\alpha_0(\Lambda)$ in
Eqs.  (\ref{alpha_lt_alphaw})-(\ref{alf0up_uw}) are analytic, it is convenient
to use them for this purpose. The resultant suggested upper bounds are of the
form
\beq
\beta(\Lambda_{dual}) < \beta_{u,w}(\Lambda_{dual}) \ , 
\label{bbdual}
\eeq
where, as above, the subscript $w$ refers to the connection with
$W(\Lambda,q)$.  In order of increasing $p$ for duals of Archimedean lattices
comprised of $p$-gons, we have
\beq
\beta_{u,w}([3 \cdot 12^2]) = [\alpha_{u,w}((3 \cdot 12^2))]^2 = 
\frac{3^{2/3} \times (2047)^{1/3}}{4} = 
6.602779477
\label{betaup_31212dual}
\eeq
\beq
\beta_{u,w}([4 \cdot 8^2]) =  [\alpha_{u,w}((4 \cdot 8^2))]^2 = 
\frac{7^{1/2} \times (127)^{1/2}}{4} = 
7.45402576
\label{betaup_488dual}
\eeq
\beq
\beta_{u,w}([4 \cdot 6 \cdot 12]) = [\alpha_{u,w}((4 \cdot 6 \cdot 12))]^2 = 
\frac{7^{1/2} \times (31)^{1/3} \times (2047)^{1/6}}{4} = 
7.403934483
\label{betaup_4612dual}
\eeq
\beq
\beta_{u,w}([6^3]) = \beta_{u,w}((3^6)) \equiv \beta_{u,w}(tri) = [\alpha_{u,w}((6^3))]^2 = 
\frac{31}{4} = 7.75 
\label{betaup_tri}
\eeq
\beq
\beta_{u,w}([3 \cdot 6 \cdot 3 \cdot 6]) = 
\alpha_{u,w}((3 \cdot 6 \cdot 3 \cdot 6)) =
\frac{3^{2/3} \times (31)^{1/3}}{2} = 
3.2671675385
\label{betaup_kagdual}
\eeq
\beq
\beta_{u,w}([3 \cdot 4 \cdot 6 \cdot 4]) = 
\alpha_{u,w}((3 \cdot 4 \cdot 6 \cdot 4)) = 
\frac{3^{1/3} \times 7^{1/2} \times (31)^{1/6}}{2} = 
3.381580457
\label{betaup_3464dual}
\eeq
\beq
\beta_{u,w}([4^4]) = \alpha_{u,w}((4^4)) \equiv \alpha_{u,w}(sq) = \frac{7}{2} 
\label{betaup_sq}
\eeq
\beq
\beta_{u,w}([3^3 \cdot 4^2]) = [\alpha_{u,w}((3^3 \cdot 4^2))]^{2/3} = 
\frac{3^{2/3} \times 7^{1/3}}{2^{2/3}} = 
2.50664897
\label{betaup_33344dual}
\eeq
\beq
\beta_{u,w}([3^2 \cdot 4 \cdot 3 \cdot 4]) = 
[\alpha_{u,w}((3^2 \cdot 4 \cdot 3 \cdot 4))]^{2/3} = 
\frac{3^{2/3} \times 7^{1/3}}{2^{2/3}} = 
2.50664897
\label{betaup_33434dual}
\eeq
\beq
\beta_{u,w}([3^4 \cdot 6]) =  
[\alpha_{u,w}((3^4 \cdot 6))]^{2/3} = 
\frac{3^{8/9} \times (31)^{1/9}}{2^{2/3}} =
2.44978501
\label{betaup_33336dual}
\eeq
and
\beq
\beta_{u,w}([3^6]) = \beta_{u,w}((6^3)) \equiv \beta_{u,w}(hc) =  
[\alpha_{u,w}((3^6))]^{1/2} = \frac{3}{\sqrt{2}} = 
2.12132034 \ . 
\label{betaup_hc}
\eeq
These values are listed in Table \ref{uw_duals_table}. 


\subsection{Connection with $W(\Lambda_{dual},q)$}

In addition to the lower bound for $W(\Lambda,q)$ on Archimedean lattices
$\Lambda$, (\ref{wwlow}) with (\ref{wlowform}), Ref. \cite{wn} also proved a
general lower bound for $W(\Lambda_{dual},q)$ on dual Archimedean lattices
$\Lambda_{dual}$.  This lower bound, applicable for $q \ge
\chi(\Lambda_{dual})$, is
\beq
W(\Lambda_{dual},q) \ge W(\Lambda_{dual},q)_\ell \ , 
\label{wwlowdual}
\eeq
where (see Eq. (5.1) of \cite{wn}) 
\beq
W \bigg ( [\prod_i \Delta_i^{b_i}], q \bigg )_\ell = 
\frac{[D_{p}(q)]^{\nu_p}}{q-1}  \ . 
\label{wlowformdual}
\eeq
As with Archimedean lattices, this naturally leads
to the conjecture that evaluating Eq. (\ref{wlowformdual}) at $q=-1$ and 
$q=0$ would yield upper bounds on $\alpha(\Lambda_{dual})$ and 
$\alpha_0(\Lambda_{dual})$, respectively, i.e., 
\beq
\alpha(\Lambda_{dual}) < \alpha_{u,w}(\Lambda_{dual})
\label{alfup_archdual}
\eeq
and
\beq
\alpha_0(\Lambda_{dual}) < \alpha_{0,u,w}(\Lambda_{dual}) \ , 
\label{alfup0_archdual}
\eeq
where
\beq
\alpha_{u,w}(\Lambda_{dual}) = \frac{|D_{p}(-1)|^{\nu_p}}{2} 
\label{alfup_archdualform}
\eeq
and
\beq
\alpha_{0,u,w}(\Lambda_{dual}) = |D_{p}(0)|^{\nu_p} \ . 
\label{alf0up_archdualform}
\eeq
The values of $\alpha_{u,w}(\Lambda_{dual})$ and
$\alpha_{0,u,w}(\Lambda_{dual})$ are listed below:
\beqs
&& \alpha_{u,w}(\Lambda_{dual}) = \frac{9}{2} \quad {\rm and} \quad 
\alpha_{0,u,w}(\Lambda_{dual}) = 4 \quad 
{\rm for} \ \Lambda_{dual} = 
[3 \cdot 12^2], \quad [4 \cdot 8^2], \quad [4 \cdot 6 \cdot 12], \quad [6^3]
\cr\cr
&& 
\label{alfup_archdual_p3}
\eeqs
\beqs
&&\alpha_{u,w}(\Lambda_{dual}) = \frac{7}{2} \quad {\rm and} \quad 
\alpha_{0,u,w}(\Lambda_{dual}) = 3 \quad {\rm for} \ \Lambda_{dual} = 
[3 \cdot 6 \cdot 3 \cdot 6], \quad [3 \cdot 4 \cdot 6 \cdot 4], \quad [4^4] 
\cr\cr
&&
\label{alfup_archdual_p4}
\eeqs
and
\beqs
&&\alpha_{u,w}(\Lambda_{dual}) = 2^{-1} \times (15)^{2/3} = 3.04110100 
\quad {\rm and} \quad 
\alpha_{0,u,w}(\Lambda_{dual}) = 2^{4/3} =2.5198421 \cr\cr
&& {\rm for} \ \Lambda_{dual} = 
[3^3 \cdot 4^2], \quad [3^2 \cdot 4 \cdot 3 \cdot 4], \quad [3^4 \cdot 6] \ . 
\label{alfup_archdual_p5}
\eeqs
As implied by duality, for the homopolygonal Archimedean lattices the 
values of $\alpha_{u,w}([3^6])=\alpha_{u,w}((6^3))=\alpha_{u,w}(hc)$ and 
$\alpha_{0,u,w}([3^6])=\alpha_{0,u,w}((6^3))=\alpha_{0,u,w}(hc)$ are the
same as those given above in Eqs. (\ref{wupperbound_alpha_hc}) and 
(\ref{wupperbound_alpha0_hc}); the values of 
$\alpha_{u,w}([4^4])=\alpha_{u,w}(sq)$ and 
$\alpha_{0,u,w}([4^4])=\alpha_{0,u,w}(sq)$ are the same as those given in 
Eqs. (\ref{wupperbound_alpha_sq}) and (\ref{wupperbound_alpha0_sq}); 
and the values of 
$\alpha_{u,w}([6^3])=\alpha_{u,w}((3^6))=\alpha_{u,w}(tri)$ and 
$\alpha_{0,u,w}([6^3])=\alpha_{0,u,w}((3^6))_{uw}=\alpha_{0,u,w}(tri)$ are the
same as those given in Eqs. (\ref{wupperbound_alpha_tri}) and 
(\ref{wupperbound_alpha0_tri}).

Using different paths for the operation of the coloring matrix, Ref. \cite{wn}
also derived more stringent lower bounds on $W(\Lambda_{dual})$ for certain
dual heteropolygonal Archimedean lattices, again applicable for $q \ge
\chi(\Lambda_{dual})$:
\beq
W([3 \cdot 12^2],q) \ge \frac{[(q-2)(q-3)]^{2/3}}{(q-1)^{1/3}} 
\label{wlow_31212_better}
\eeq
\beq
W([4 \cdot 8^2],q) \ge \Bigg [ \frac{(q-2)(q^2-5q+7)}{q-1} \Bigg ]^{1/2} 
\label{wlow_488_better}
\eeq
and
\beq
W([4 \cdot 6 \cdot 12],q) \ge \frac{(q-2)(q^2-5q+7)^{1/3}}{(q-1)^{2/3}} \ . 
\label{wlow_4612_better}
\eeq
Following the same procedure as explained above, we can use these to obtain
more restrictive conjectured upper bounds on $\xi(\Lambda_{dual})$ for 
$\xi=\alpha, \ \alpha_0$ and $\Lambda_{dual}=[3 \cdot 12^2], \ [4 \cdot 8^2]$,
and $[4 \cdot 6 \cdot 12]$. These are marked with primes to
distinguish them from the upper bounds given above: 
\beq
\alpha_{u,w'}([3 \cdot 12^2]) = 2 \times 3^{2/3} = 4.160168
\label{wwupperbound_alpha_31212dual}
\eeq
\beq
\alpha_{0,u,w'}([3 \cdot 12^2]) = 6^{2/3} = 3.301927
\label{wwupperbound_alpha0_31212dual}
\eeq
\beq
\alpha_{u,w'}([4 \cdot 8^2]) = \sqrt{\frac{39}{2}} = 4.415880 
\label{wwupperbound_alpha_488dual}
\eeq
\beq
\alpha_{0,u,w'}([4 \cdot 8^2]) = \sqrt{14} = 3.741657 
\label{wwupperbound_alpha0_488dual}
\eeq
\beq
\alpha_{u,w'}([4 \cdot 6 \cdot 12]) = 2^{-2/3} \times 3 \times 13^{1/3} = 
4.443744
\label{wwupperbound_alpha_4612dual}
\eeq
and
\beq
\alpha_{0,u,w'}([4 \cdot 6 \cdot 12]) = 2 \times 7^{1/3} = 3.825862 \ .
\label{wwupperbound_alpha0_4612dual}
\eeq
The values of $\alpha_{u,w}(\Lambda_{dual})$ and 
$\alpha_{0,u,w}(\Lambda_{dual})$, or the more restrictive values 
$\alpha_{u,w'}(\Lambda_{dual})$ and $\alpha_{0,u,w'}(\Lambda_{dual})$ where
they apply, are listed in Table \ref{uw_duals_table}. 

We recall that the lower bounds that were proved to apply for $W(\Lambda,q)$
for $q \ge \chi(\Lambda)$ on all Archimedean lattices $\Lambda$ in \cite{wn}
were found to be very close to the actual values of $W(\Lambda,q)$ as
determined by Monte-Carlo simulations and series expansions (and by the exact
result for $W(tri)$ in \cite{baxter87}). This led to the expectation that,
not only would (\ref{alfup_uw}) and (\ref{alf0up_uw}) constitute 
upper bounds on $\alpha(\Lambda)$ and $\alpha_0(\Lambda)$, respectively (which
our results support), but also that the values of 
$\alpha_{u,w}(\Lambda)$ and $\alpha_{0,u,w}(\Lambda)$ would be close
to the actual values of $\alpha(\Lambda)$ and $\alpha_0(\Lambda)$.  This
expectation was confirmed for homopolygonal Archimedean lattices in \cite{ac}.
Here, we also confirm this for the heteropolygonal Archimedean lattices for
which we have obtained approximate values of these exponential growth
constants, namely for the $(4 \cdot 8^2)$, $(3 \cdot 6 \cdot 3 \cdot 6)$, $(3^3
\times 4^2)$, and $(3^2 \cdot 4 \cdot 3 \cdot 4)$ lattices.

Furthermore, by duality, one can calculate values of $\beta_{u,w}(\Lambda)$ on
Archimedean lattices corresponding to these values of
$\alpha_{u,w}(\Lambda_{dual})$ on duals of Archimedean lattices.  We obtain
\beqs
&&\beta_{u,w}(\Lambda) = [\alpha_{u,w}(\Lambda_{dual})]^{1/2} = \frac{3}{\sqrt{2}} =
2.121320 \cr\cr
&& {\rm for} \ \Lambda = (3 \cdot 12^2), \quad (4 \cdot 8^2), \quad 
(4 \cdot 6 \cdot 12), \quad (6^3) 
\label{beta_lam_delta3}
\eeqs
\beqs
&&\beta_{u,w}(\Lambda) = \alpha_{u,w}(\Lambda_{dual}) = \frac{7}{2} \cr\cr
&& {\rm for} \ \Lambda = (3 \cdot 6 \cdot 3 \cdot 6), \quad (3 \cdot 6 \cdot 4 \cdot 6), \quad 
(4^4) 
\label{beta_lam_delta4}
\eeqs
\beqs
&&\beta_{u,w}(\Lambda) = [\alpha_{u,w}(\Lambda_{dual})]^{3/2} = \frac{15}{2\sqrt{2}} =
5.303301 \cr\cr
&& {\rm for} \ \Lambda = (3^3 \cdot 4^2), \quad (3^2 \cdot 4 \cdot 3 \cdot 4), \quad 
(3^4 \cdot 6) 
\label{beta_lam_delta5}
\eeqs
and
\beq
\beta_{u,w}(3^6) \equiv \beta_{u,w}(tri) = [\alpha_{u,w}(hc)]^2 = \frac{31}{4}
= 7.75 \ . 
\label{beta_lam_delta6}
\eeq
Using duality together with the values of $\alpha_{u,w'}([3 \cdot 12^2])$, 
$\alpha_{u,w'}([4 \cdot 8^2])$, and $\alpha_{u,w'}([4 \cdot 6 \cdot 12])$,
we obtain the more restrictive conjectured upper limits 
\beq
\beta_{u,w'}((3 \cdot 12^2)) = [\alpha_{u,w'}([3 \cdot 12^2])]^{1/2} = 
2^{1/2} \times 3^{1/3} = 2.039649
\label{wwupperbound_beta_31212}
\eeq
\beq
\beta_{u,w'}((4 \cdot 8^2)) = [\alpha_{u,w'}([4 \cdot 8^2])]^{1/2} = 
\bigg ( \frac{39}{2} \bigg )^{1/4} = 2.101400 
\label{wwupperbound_beta_488}
\eeq
and
\beq
\beta_{u,w'}((4 \cdot 6 \cdot 12)) = 
[\alpha_{u,w'}([4 \cdot 6 \cdot 12])]^{1/2} = 
2^{-1/3} \times 3^{1/2} \times (13)^{1/16} = 2.108019 \ . 
\label{wwupperbound_beta_4612}
\eeq
We list these values of $\beta_{u,w}(\Lambda)$ or $\beta_{u,w'}(\Lambda)$ in
Table \ref{uw_table}. Since the application of the duality transformation twice
is the identity map, it follows that for the homopolygonal Archimedean
lattices, the values of $\beta_{u,w}(hc)$, $\beta_{u,w}(sq)$, and
$\beta_{u,w}(tri)$ in Eqs.  (\ref{beta_lam_delta3}), (\ref{beta_lam_delta4}),
and (\ref{beta_lam_delta6}) are equal to the values obtained in
Eqs. (\ref{betaup_hc}), (\ref{betaup_sq}), and (\ref{betaup_tri}).

We observe that where we can compare the values of $\beta_{u,w}(\Lambda)$ or
$\beta_{u,w'}(\Lambda)$ with the values $\beta_{ap}(\Lambda)$ that we have
determined above via calculations with infinite-length, finite-width strips,
namely for the $(4 \cdot 8^2)$, $(3 \cdot 6 \cdot 3 \cdot 6)$, $(3^3 \times
4^2)$, and $(3^2 \cdot 4 \cdot 3 \cdot 4 \cdot 3)$ lattices, they are
reasonably close for each $\Lambda$. We will use this finding
below. Interestingly, for the $(3^3 \cdot 4^2)$ and $(3^2 \cdot 4 \cdot 3 \cdot
4)$ lattices, the common value of $\beta_{u,w}((3^3 \cdot
4^2))=\beta_{u,w}((3^2 \cdot 4 \cdot 3 \cdot 4)) = 5.303301$ in
Eq. (\ref{beta_lam_delta5}) lies slightly below the upper bounds that we infer
for these lattices from our computations with infinite-length, finite-width
strips.


\section{Comparative Analysis}
\label{comparison_section}

With these calculations on heteropolygonal Archimedean lattices, we extend our
results in \cite{ac} for homopolygonal Archimedean lattices to the full set of
Archimedean lattices.  We find that for all Archimedean lattices, the values of
$\alpha(\Lambda)$, $\alpha_0(\Lambda)$, and $\beta(\Lambda)$ that are
consistent with our inferred upper and lower bounds, and the exact values where
we have calculated them, are monotonically increasing functions of
$\Delta(\Lambda)$.  In particular, this applies to the average values,
$\alpha_{ave}(\Lambda)$, $\alpha_{0,ave}(\Lambda)$, and $\beta_{ave}(\Lambda)$
and to the $\alpha_{u,w}(\Lambda)$, $\alpha_{0,u,w}(\Lambda)$, and
$\beta_{u,w}(\Lambda)$ values. These statements are also true of our results
for the dual Archimedean lattices $\Lambda_{dual}$.  We recall that for a
$\Delta$-regular graph $G$, the number of edges is related to the number of
vertices by $e(G)=(\Delta/2) n$, so the exponential growth constant
$\epsilon(\{G\})$ increases with $\Delta$, as
$\epsilon(\{G\})=2^{\Delta/2}$. Thus, the monotonic increase that we find for
$\alpha(\Lambda)$, $\alpha_0(\Lambda)$, and $\beta(\Lambda)$ as functions of
$\Delta(\Lambda)$ on these lattices can be interpreted as a consequence of the
fact that, for a section $G_\Lambda$ of the lattice $\Lambda$ with
$n(G_\Lambda) \to \infty$, an increase in $\Delta(G_\Lambda)$ leads, via the
exponential increase in $N_{eo}(G_\Lambda)$, to a commensurately large
exponential increase in $a(G_\Lambda)$, $a_0(G_\Lambda)$, and $b(G_\Lambda)$.

In \cite{ac} we observed that the increase in these exponential growth
constants with vertex degree $\Delta(\Lambda)$ for the homopolygonal
Archimedean lattices is the opposite of what was found for the the behavior of
$W(\Lambda,q)$ for these lattices $\Lambda$ with $q$ in the range $q \ge
\chi(\Lambda)$ used for proper $q$-coloring. In \cite{w} (see, e.g., Fig. 5),
$W(\Lambda,q)$ was shown to be a monotonically decreasing function of
$\Delta(\Lambda)$ for $q \ge \chi(\Lambda)$. This dependence was also shown for
the upper and lower bounds on $W(\Lambda,q)$ for Archimedean lattices,
including heteropolygonal lattices, in \cite{wn,ww,w3} (see also
\cite{w,ilb}). This is a consequence of the property that an increase in
$\Delta(\Lambda)$ generically increases the constraints on a proper
$q$-coloring of the lattice $\Lambda$ \cite{w,bcc99,wn}.  The reversal in the
dependence of $W(\Lambda,q)$ on $\Delta(\Lambda)$ when one switches from $q \ge
\chi(\Lambda)$ to $q \le 0$ was seen in (Fig. 5 of) Ref. \cite{w}. Here we have
extended this contrast from the homopolygonal lattices studied in \cite{ac} to
heteropolygonal Archimedean lattices, and hence to all Archimedean lattices.

As we did for homopolygonal Archimedean lattices in \cite{ac}, we next compare
our results for $\alpha(\Lambda)$, $\alpha_0(\Lambda)$, and $\beta(\Lambda)$
with the exponential growth constants for spanning trees on heteropolygonal
Archimedean lattices. We first review the relevant definitions.  A tree graph
is a connected graph that does not contain any circuits, and a spanning tree of
a graph $G$ is a subgraph of $G$ that is a tree and that contains all of the
vertices of $G$.  From Eq. (\ref{t}), it follows that the number of spanning
trees in a graph $G$, denoted $N_{ST}(G)$, is given by the following evaluation
of the Tutte polynomial:
\beq
N_{ST}(G)=T(G,1,1) \ .
\label{nst}
\eeq
A different way to calculate $N_{ST}(G)$, which has been the basis of a number
of exact calculations, starts with the adjacency matrix $A$ of the graph
$G$. Let us denote the number of edges connecting two adjacent vertices $v_i$
and $v_j$ as $N(e_{ij})$. The adjacency matrix $A$ is an $n(G) \times n(G)$
matrix with elements $A_{ij}=N(e_{ij})$ if the vertices $v_i$ and $v_j$ are
adjacent and $A_{ij}=0$ otherwise. The Laplacian matrix $Q$ is an $n(G) \times
n(G)$ matrix with elements $Q_{ij} = \Delta(G)\delta_{ij} - A_{ij}$, where
$\delta_{ij}$ is the Kronecker delta function. The sum of the elements in any
row or column of $Q$ is zero, and consequently, one of the eigenvalues of $Q$
is zero.  Denote the remaining eigenvalues as
$\lambda^{(Q)}_1,...,\lambda^{(Q)}_{n(G)-1}$. Then \cite{graphtheory}
\beq
N_{ST}(G) = \frac{1}{n(G)} \, \prod_{s=1}^{n(G)-1} \lambda^{(Q)}_s \ . 
\label{nstdet}
\eeq

For the lattice graphs $G$ studied here, $N_{ST}(G)$ grows
exponentially rapidly with the number of vertices, $n(G)$.  It is then natural
to define the corresponding exponential growth constant,
\beq
\tau(\{G\}) = \lim_{n(G) \to \infty} [N_{ST}(G)]^{1/n(G)} \ .
\label{tau}
\eeq
An equivalent quantity that has often been used in previous works is $z(\{G\})
= \ln[\tau(\{ G \})]$. For a lattice graph, in the limit $n(G) \to \infty$, the
logarithm of the product of eigenvalues in Eq. (\ref{nstdet}), which gives
$z(\Lambda)$, becomes an integral, whose integrand is determined from a
knowledge of the basis vectors of the lattice (see Eq. (4.16) in
\cite{st}). This integral formulation has been used for the exact calculation
of $z(\Lambda)$, or equivalently, $\tau(\Lambda)$ for all of the Archimedean
lattices.  Specifically, $z(\Lambda)$ was calculated for the square,
triangular, and honeycomb lattices in \cite{wu77}; for the $(3 \cdot 12^2)$ and
$(3 \cdot 6 \cdot 3 \cdot 6)$ lattices in \cite{st}; for the $(4 \cdot 8^2)$
lattice in \cite{st,std}; and for the $(4 \cdot 6 \cdot 12)$, $(3 \cdot 4 \cdot
6 \cdot 4)$, $(3^3 \cdot 4^2)$, $(3^2 \cdot 4 \cdot 3 \cdot 4)$, and $(3^4
\cdot 6)$ lattice in \cite{sti}.  We list the numerical values of
$\tau(\Lambda)$ for these lattices in Table \ref{uw_table}.  By duality, the
values of $\tau(\Lambda_{dual})$ on the dual Archimedean lattices are exactly
determined in terms of these $\tau(\Lambda)$ values. We list the numerical
values of $\tau(\Lambda_{dual})$ in Table \ref{uw_duals_table}. 

A theorem of Thomassen \cite{thomassen2010} states that if $G$ is a
$\Delta$-regular graph of degree $\Delta(G) \le 3$ which has no loops (but
which may have bridges and multiple edges), then $N_{ST}(G) \le a(G)$.
Considering a family of graphs of this type and taking the limit $n(G) \to
\infty$, this implies that in this limit, $\tau(\{G\}) \le \alpha(\{ G \})$.
As must be true, in agreement with this theorem (for the special case relevant
to our application to graphs without multiple edges), our result for the
approximate value of $\alpha((4 \cdot 8^2))$ in Eq. (\ref{alpha_est_488}) is
greater than the exact result for $\tau((4 \cdot 8^2))$ from \cite{st,std}, as
listed in Table \ref{egc_values_table}. The values of $\alpha_{u,w}(\Lambda)$
for all of the $\Delta(\Lambda)=3$ Archimedean lattices are also in agreement
with this theorem, as is evident from Table \ref{uw_table}.

In \cite{ac} we observed that our determinations of $\alpha(\Lambda)$ and
$\beta(\Lambda)$ on the homopolygonal Archimedean lattices $\Lambda=hc, \ sq, \
tri$ were in agreement with an inequality on exponential growth constants
implied by the Merino-Welsh conjecture \cite{merino_welsh99}.  Here we extend
our investigation of this subject to the set of all Archimedean lattices.  We
first recall the Merino-Welsh conjecture. Let $G$ be a connected graph without
loops or bridges (which may have multiple edges, although we restrict here to
graphs without multiple edges). Then the Merino-Welsh
conjecture (MWC) is the inequality \cite{merino_welsh99}
\beq
N_{ST}(G) \le {\rm max}[a(G), \ b(G)] \ , \quad i.e., \
T(G,1,1) \le {\rm max}[T(G,2,0), \ T(G,0,2)] \quad {\rm (MWC)} \ .
\label{mwc}
\eeq
In the later paper \cite{conde_merino2009}, Conde and Merino conjectured the
stronger inequality that if $G$ is a connected graph without loops or bridges
(which may have multiple edges), then
\beq
[N_{ST}(G)]^2 \le a(G)b(G) \ , \quad i.e., \
[T(G,1,1)]^2 \le T(G,2,0)T(G,0,2) \quad {\rm (CMC)} \ ,
\label{cmc}
\eeq
where our abbreviation CMC stands for Conde-Merino conjecture. 
Some relevant related papers include \cite{merino_note2009}-\cite{knauer2016}.
For our purposes, we first observe that the Merino-Welsh and Conde-Merino 
conjectures imply the following inequalities on exponential growth constants,
where $\{G\}$ is the $n(G) \to \infty$ limit of graphs $G$ that satisfy the
premise of the MWC and CMC: 
\beq
\tau(\{ G \}) \le {\rm max}[ \alpha(\{ G \}), \ \beta( \{ G \})] \quad
{\rm from \ MWC} \ .
\label{mwc_egc}
\eeq
and
\beq
[\tau(\{ G \})]^2 \le \alpha(\{ G \})\beta( \{ G \}) \quad {\rm from \ CMC}
\ .
\label{cmc_egc}
\eeq
For the comparison, we make use of the approximate values
$\alpha_{ap}(\Lambda)$, $\alpha_{0,ap}(\Lambda)$, and $\beta_{ap}(\Lambda)$
that we have determined from our upper and lower bounds for the $(4 \cdot
8^2)$, $(3 \cdot 6 \cdot 3 \cdot 6)$, $(3^3 \cdot 4^2)$, and $(3^2 \cdot 4
\cdot 3 \cdot 4)$ Archimedean lattices.  As is evident in Table
\ref{egc_values_table}, these are in agreement with the inequalities
(\ref{mwc_egc}) and (\ref{cmc_egc}) implied, respectively, by the Merino-Welsh
conjecture and the Conde-Merino conjecture.  Our results in \cite{ac} and here
are also useful as a quantitative measure of how close to being saturated the
inequalities (\ref{mwc_egc}) and (\ref{cmc_egc}) are. 

If we assume that for the other heteropolygonal Archimedean lattices, our
values of $\alpha_{u,w}(\Lambda)$, $\beta_{u,w}(\Lambda)$, and
$\beta_{u,w'}(\Lambda)$ are close to the actual respective values, then we can
substitute these into (\ref{mwc_egc}) together with the known values of
$\tau(\Lambda)$ for comparisons.  As is evident in Table \ref{uw_table}, in all
cases, this comparison agrees with the inequalities (\ref{mwc_egc}) and 
(\ref{cmc_egc}) implied by the Merino-Welsh and Conde-Merino conjectures.  
This is also true of our results for the dual
Archimedean lattices $\Lambda_{dual}$, as one can see from Table
\ref{uw_duals_table}.

Moreover, with the same assumptions as above, we can combine our calculations
on Archimedean lattices to comment on the relative sizes of $\alpha(\Lambda)$,
$\beta(\Lambda)$, and $\tau(\Lambda)$ as functions of $\Delta(\Lambda)$. We
find
\beq
\alpha(\Lambda) > \tau(\Lambda) > \beta(\Lambda) \quad {\rm for} \
\Delta(\Lambda)=3
\label{abt_order_deg3_arch}
\eeq
\beq
\alpha(\Lambda), \ \beta(\Lambda) > \tau(\Lambda) \quad {\rm for} \
\Delta(\Lambda)=4
\label{abt_order_deg4_arch}
\eeq
and
\beq
\beta(\Lambda) > \tau(\Lambda) > \alpha(\Lambda) \quad {\rm for} \
\Delta(\Lambda)=5, \ 6
\label{abt_order_deg56_arch}
\eeq
Although the duals of heteropolygonal Archimedean lattice are not
$\Delta$-regular, i.e., have vertices of different degrees, one can explore the
dependence of these exponential growth constants on the effective vertex degree
$\Delta_{eff}(\Lambda_{dual})$ given in Eq. (\ref{deltaeff_lambdadual}). With
the same assumptions as above, we find similar inequalities for
$\alpha_{u,w}(\Lambda_{dual})$, $\beta_{u,w}(\Lambda_{dual})$ and the exactly
known $\tau(\Lambda_{dual})$ values (including also our results on
homopolygonal lattices).
\beq
\alpha(\Lambda_{dual}) > \tau(\Lambda_{dual}) > \beta(\Lambda_{dual}) 
\quad {\rm for} \ \Delta_{eff}(\Lambda_{dual})=3, \ \frac{10}{3} \ , 
\label{abt_order_deg3and3p3_archdual}
\eeq
\beq
\alpha(\Lambda_{dual}), \ \beta(\Lambda_{dual}) > \tau(\Lambda_{dual}) 
\quad {\rm for} \ \Delta_{eff}(\Lambda_{dual})=4
\label{abt_order_deg4_archdual}
\eeq
and
\beq
\beta(\Lambda_{dual}) > \tau(\Lambda_{dual}) > \alpha(\Lambda_{dual}) 
\quad {\rm for} \ \Delta_{eff}(\Lambda_{dual})=6 \ . 
\label{abt_order_deg6_archdual}
\eeq
%


\section{Conclusions}
\label{conclusion_section}

In this paper, extending our study in \cite{ac}, we have inferred upper and
lower bounds on the exponential growth constants $\alpha(\Lambda)$,
$\alpha_0(\Lambda)$, and $\beta(\Lambda)$ that characterize the asymptotic
behavior of acyclic orientations, acyclic orientations with a single source
vertex, and totally cyclic orientations of heteropolygonal Archimedean
lattices. To our knowledge, these are the best bounds on these quantities. As
in the case of the homopolygonal Archimedean lattices (honeycomb, square, and
triangular), these bounds converge quickly, even for moderate values of $L_y$,
the strip width. Furthermore, again as with the homopolygonal lattices, the
upper and lower bounds are close to each other, which enables us to infer
approximate values of the actual exponential growth constants themselves.  A
general property that we observe is that $\alpha(\Lambda)$,
$\alpha_0(\Lambda)$, and $\beta(\Lambda)$ are monotonically increasing
functions of vertex degree $\Delta(\Lambda)$ for all Archimedean lattices, both
homopolygonal and heteropolygonal.  We have conjectured that analytic
expressions that were proved in \cite{wn} to be lower bounds on $W(\Lambda,q)$
for values of $q$ used in proper $q$-colorings of Archimedean and dual
Archimedean lattices $\Lambda$ and $\Lambda_{dual}$ provide upper bounds on
$\alpha(\Lambda)$, $\alpha_0(\Lambda)$, $\alpha(\Lambda_{dual})$, and
$\alpha_0(\Lambda_{dual})$. We have also used duality relations to obtain
corresponding conjectured upper bounds on $\beta(\Lambda)$ and
$\beta(\Lambda_{dual})$. In all cases, these are consistent with the upper
bounds that we derive from our calculations using infinite-length, finite-width
lattice strips of these graphs.  We have also made comparisons with the
exponential growth constants for spanning trees on these lattices, finding
agreement with inequalities that follow from the Merino-Welsh and Conde-Merino
conjectures. In addition to providing support for these inequalities, our
results give a quantitative measure of how close to being saturated they are 
for the lattices that we study.


\begin{acknowledgments}

This research was supported in part by the Taiwan Ministry of Science and
Technology grant MOST 103-2918-I-006-016 (S.-C.C.) and by
the U.S. National Science Foundation grant No. NSF-PHY-16-1620628 (R.S.).

\end{acknowledgments}


\begin{appendix}


\section{Some Graph Theory Background}
\label{graphtheory}

In this appendix we briefly list some formulas that are relevant to our
analysis in the text. As stated in the text, we denote a graph as $G=(V,E)$
with vertex and edge sets $V$ and $E$. We let $n=n(G)=|V|$, $e(G)=|E|$,
$fc(G)$, and $k(G)$ denote the number of vertices, edges, faces, and connected
components of $G$, respectively. The degree of a vertex in a graph is the
number of edges that connect to it. Graphs whose vertices all have the same
degree $\Delta$ is called a $\Delta$-regular graph.  The girth $g(G)$ of a
graph is the length of edges in a minimal-distance circuit on $G$.  (If $G$ has
no circuits, then $g(G)$ is not defined.)

An Archimedean lattice is a tiling of the (infinite) plane with one or more
types of regular polygons (i.e., polygons whose sides all have equal length and
whose internal angles are all equal) such that all vertices are equivalent. As
discussed in the text, this means that an Archimedean lattice $\Lambda$ can be
defined as the ordered sequence of polygons that one traverses in a circuit
around any vertex, $\Lambda = \prod_i p_i^{a_i}$, where the $i$'th polygon has
$p_i$ sides and appears $a_i$ times together in the sequence.  There are eleven
Archimedean lattices. Of these, three are homopolygonal, namely $(4^4)$
(square), $(3^6)$ (triangular), and $(6^3)$ (honeycomb), and the other eight
are heteropolygonal. Synonymous notations include $(3 \cdot 12^2) \equiv (3
\cdot 12 \cdot 12)$, $(4 \cdot 8^2) \equiv (4 \cdot 8 \cdot 8)$, etc.  The
Archimedean lattices, listed in order of increasing vertex degree
$\Delta(\Lambda)$, and, for a given $\Delta(\Lambda)$ in order of increasing
girth $g(\Lambda)$, are
\beq
\Delta(\Lambda)=3: \quad (3 \cdot 12^2), \quad (4 \cdot 8)^2, \quad 
(4 \cdot 6 \cdot 12), \quad (6)^3
\label{arch_degree3}
\eeq
\beq
\Delta(\Lambda)=4: \quad (3 \cdot 6 \cdot 3 \cdot 6), \quad
(3 \cdot 4 \cdot 6 \cdot 4) \ , \quad (4^4)
\label{arch_degree4}
\eeq
\beq
\Delta(\Lambda)=5: \quad (3^3 \cdot 4^2), \quad (3^2 \cdot 4 \cdot 3 \cdot 4),
\quad (3^4 \cdot 6) \ ,
\label{arch_degree5}
\eeq
and
\beq
\Delta(\Lambda)=6: \quad (3^6)
\label{arch_degree6}
\eeq

Let $G$ be a planar graph, indicated as $G_{pl}$, and denote $G_{pl}^*$ as the
(planar) dual graph. Then the Tutte polynomial satisfies $T(G_{pl},x,y) =
T(G_{pl}^*,y,x)$. Consequently, $a(G_{pl}) = b(G_{pl}^*)$.  From duality, one
has the equality $n(G_{pl}^*)=fc(G_{pl})$.  Recall the Euler relation that for
a planar graph $G_{pl}$, $fc(G_{pl})-e(G_{pl})+n(G_{pl})=2$. Following the
notation in \cite{st}, for a $\Delta$-regular planar graph $G_{pl}$, we define
the ratio
\beq
\nu_{ \{ G_{pl} \} }\equiv \lim_{n(G)\to \infty}\frac{n(G_{pl}^*)}{n(G_{pl})}=
\lim_{n(G)\to \infty}\frac{fc(G_{pl})}{n(G_{pl})} \ .
\label{nu_g}
\eeq
Using the Euler relation and the fact that $\Delta(G)=2e(G)/n(G)$, we have 
\beq
\nu_{ \{ G_{pl} \} } = \frac{\Delta(G_{pl})}{2}-1 \ .
\label{nuform}
\eeq
If the vertices of $G_{pl}^*$ have uniform degree, then 
$\nu(G_{pl}^*) = 1/\nu(G_{pl})$.  In general, even if the vertices of 
$G_{pl}^*$ do not have uniform degree, in the limit $n(G) \to \infty$,
\beq
\nu(\{ G_{pl}^* \}) = \frac{1}{\nu(\{ G \})} \ . 
\label{nunudual}
\eeq

For homopolygonal lattices,
\beq
\nu(sq) = 1, \quad \quad \nu(hc)= \frac{1}{\nu(tri)} = \frac{1}{2} \ . 
\label{nuhomo}
\eeq
For heteropolygonal lattices, we have 
\beq
\nu((3 \cdot 12^2)) = \nu((4 \cdot 8^2)) = 
\nu((4 \cdot 6 \cdot 12)) = \frac{1}{2}
\label{nu_hetero_deg3}
\eeq
\beq
\nu((3 \cdot 6 \cdot 3 \cdot 6)) = \nu((3 \cdot 4 \cdot 6 \cdot 4) = 1
\label{nu_hetero_deg4}
\eeq
and
\beq
\nu((3^3 \cdot 4^2)) = \nu((3^2 \cdot 4 \cdot 3 \cdot 4)) = \nu((3^4 \cdot 6))
= \frac{3}{2} \ . 
\label{nu_hetero_deg5}
\eeq

For a $\Delta$-regular planar graph $G_{pl}$, in the $n(G_{pl}) \to \infty$
limit, 
\beqs 
\beta(\{ G_{pl} \}) &=& 
\lim_{n(G_{pl}) \to \infty} [T(G_{pl},0,2)]^{\frac{1}{n(G_{pl})}} = 
\lim_{n(G_{pl}^*) \to \infty} 
[T(G_{pl}^*,2,0)]^{\frac{\nu(G_{pl})}{n(G_{pl}^*)}} \cr\cr
&=& [\alpha(\{ G_{pl}^* \})]^{\nu(G_{pl})} \ . 
\label{beta_alphadual}
\eeqs
Similarly, 
\beq 
\alpha(\{ G_{pl} \})) = [\beta(\{ G_{pl}^* \})]^{\nu(G_{pl})} \ , \ i.e., \quad
 \beta(\{ G_{pl}^* \}) =[\alpha(\{ G_{pl} \} )]^{1/\nu(G_{pl})}
\label{alpha_betadual}
\eeq
In \cite{ac} we appliled these relations to determine $\beta(hc)$ in terms of
$\alpha(tri)$, using
\beq 
\beta(hc)=[\alpha(tri)]^{\nu(hc)} = [\alpha(tri)]^{1/2}
\label{betahc_sqrt_alpha_tri}
\eeq

Here we will use these relations to determine
$\beta(\Lambda_{dual})$ for the duals of heteropolygonal lattices. In general,
for each of the duals of Archimedean lattices, $\Lambda_{dual}$, we have 
\beq
\beta(\Lambda_{dual}) = [\alpha(\Lambda)]^{\frac{1}{\nu(\Lambda)}} \ . 
\label{beta_dual}
\eeq
Specifically, 
\beq
\beta(\Lambda_{dual}) = [\alpha(\Lambda)]^2 \quad {\rm for} \ \ 
\Lambda = (3 \cdot 12^2), \quad (4 \cdot 8^2), \quad (4 \cdot 6 \cdot 12) 
\label{beta_dualdeg3}
\eeq
\beq
\beta(\Lambda_{dual})) = \alpha(\Lambda) \quad {\rm for} \ \ 
\Lambda = (3 \cdot 6 \cdot 3 \cdot 6), \quad (3 \cdot 4 \cdot 6 \cdot 4)
\label{beta_dualdeg4}
\eeq
and
\beq
\beta(\Lambda_{dual}) = [\alpha(\Lambda)]^{2/3} \quad {\rm for} \ \ 
\Lambda = (3^3 \cdot 4^2), \quad (3^2 \cdot 4 \cdot 3 \cdot 4), \quad 
(3^4 \cdot 6)  \ . 
\label{beta_dualdual5}
\eeq

Given a graph $G$, a spanning subgraph of $G$, denoted $G'$, is a graph with
the same vertex set $V$ and a subset of the edge set $E$, i.e., $G'=G'(V,E')$
with $E' \subseteq E$.  A cycle (circuit) on $G$ is defined as a set of edges
that form a closed circuit (cycle). Let $c(G)$ denote the number of linearly
independent cycles in $G$.  A tree graph is a connected graph that contains no
cycles. A spanning tree is a spanning subgraph that is a tree graph.

The chromatic polynomial of $G$, $P(G,q)$, counts the number of ways of
assigning $q$ colors to the vertices of $G$ subject to the condition that no
two adjacent vertices have the same color. This has an expression as a sum of
contributions from spanning subgraphs $G' \subseteq G$ as 
\beq
P(G,q) = \sum_{G' \subseteq G} (-1)^{e(G')} q^{k(G')} \ . 
\label{p}
\eeq
From Eq. (\ref{p}), it is clear that $P(G,q)$ always contains a factor of
$q$, so one can extract this and define a reduced polynomial 
\beq
P_r(G,q) \equiv \frac{P(G,q)}{q} \ .
\label{pr}
\eeq

The partition function of the $q$-state Potts model, $Z(G,q,v)$, has an
expression as a sum of contributions from spanning subgraphs 
$G' \subseteq G$ as \cite{fk,wurev}
\beq
Z(G,q,v) = \sum_{G' \subseteq G} v^{e(G')} q^{k(G')} \ . 
\label{z}
\eeq
The chromatic polynomial is a special case of this partition function:
$P(G,q)=Z(G,q,-1)$, where $v=-1$ corresponds to the zero-temperature limit of
the antiferromagnet. The ground-state degeneracy, per vertex, of the 
Potts antiferromagnet on a graph $G$ in the limit $n(G) \to \infty$ is 
\beq
W(\{G\},q) = \lim_{n(G) \to \infty} [P(G,q)]^{1/n(G)} \ . 
\label{w}
\eeq

The Tutte polynomial $T(G,x,y)$ is given by 

\beq
T(G,x,y) = \sum_{G' \subseteq G} (x-1)^{k(G')-k(G)} \, (y-1)^{c(G')} \ . 
\label{t}
\eeq
This is equivalent to the Potts model partition function: 
\beq
Z(G,q,v) = (x-1)^{k(G)}(y-1)^{n(G)}T(G,x,y) 
\label{zt}
\eeq
with the definitions $x = 1 + (q/v)$ and $y=v+1$. The dimensionless reduced
free energy (per vertex) of the Potts model on a graph $G$, in the limit $n(G)
\to \infty$, is
\beq 
f(\{ G\},q,v) = \lim_{n(G) \to \infty} \frac{1}{n(G)} \ln [Z(G,q,v)]
\label{f}
\eeq
The number of spanning trees on a graph $G$, denoted $N_{ST}(G)$, is given by
\beq
N_{ST}(G)=T(G,1,1) \ . 
\label{nst_tx1y1}
\eeq
With $G_{pl}$ a planar graph, one has
$N_{ST}(G_{pl})=N_{ST}(G_{pl}^*)$. Defining $\tau( \{G \}) = \lim_{n(G) \to
  \infty} [N_{ST}(G)]^{1/n(G)}$ as in the text, we have
\beq
\tau(G_{pl}^*)=[\tau(G_{pl})]^{\nu(G_{pl}^*)}=[\tau(G_{pl})]^{1/\nu(G_{pl})} 
\ . 
\label{taudual}
\eeq
Hence, for homopolygonal lattices, $\tau(hc)=[\tau(tri)]^{1/2}$ and for
heteropolygonal lattices, 
\beq
\tau(\Lambda_{dual}) = [\tau(\Lambda)]^2  \quad {\rm for} \ \ 
\Lambda = (3 \cdot 12^2), \quad (4 \cdot 8^2), \quad (4 \cdot 6 \cdot 12) 
\label{tau_dualdeg3}
\eeq
\beq
\tau(\Lambda_{dual})) = \tau(\Lambda) \quad {\rm for} \ \ 
\Lambda = (3 \cdot 6 \cdot 3 \cdot 6), \quad (3 \cdot 4 \cdot 6 \cdot 4)
\label{tau_dualdeg4}
\eeq
and
\beq
\tau(\Lambda_{dual}) = [\tau(\Lambda)]^{2/3} \quad {\rm for} \ \ 
\Lambda = (3^3 \cdot 4^2), \quad (3^2 \cdot 4 \cdot 3 \cdot 4), \quad 
(3^4 \cdot 6)  \ . 
\label{tau_dualdual5}
\eeq
Since the values of $\tau(\Lambda)$ are known exactly for all of the
Archimedean lattices, these relations yield the values of
$\tau(\Lambda_{dual})$ for all of the dual Archimedean lattices.  These are
listed in Table \ref{uw_duals_table}. 

\end{appendix}



\newpage


\begin{table}
  \caption{\footnotesize{Lower bounds and their ratios for 
$\alpha((4 \cdot 8^2))$ as functions of strip width $L_y$. In this table and 
the others, the abbreviation cyl stands for ``cylindrical''.}} 
\begin{center}
\begin{tabular}{||l|l|l|l||}
\hline
BC  & $L_y$ & $[\lambda_{(4 \cdot 8^2),L_y,free/cyl}(-1)]^{1/(4L_y)}$ & 
$R_{(4 \cdot 8^2),\frac{L_y+1}{L_y}/\frac{L_y+2}{L_y},free/cyl}(-1)$ \\ 
\hline\hline
free  & 2 & $(889)^{1/8}=2.33675252$ & \\ \hline
free  & 3 & 2.461131465              & 1.05322726 \\ \hline
free  & 4 & 2.52577995               & 1.02626779 \\ \hline
free  & 5 & 2.56538118               & 1.01567881 \\ \hline
free  & 6 & 2.592126335              & 1.01042541 \\ \hline\hline
cyl   & 2 & $\sqrt{7}=2.64575131$    &            \\ \hline
cyl   & 4 & 2.725822615              & 1.03026411 \\ \hline
cyl   & 6 & 2.7297041765             & 1.001423996 \\ \hline
\end{tabular}
\end{center}
\label{lowerbounds_alpha_488_table}
\end{table}


\begin{table}
  \caption{\footnotesize{Upper bounds and their ratios for 
$\alpha((4 \cdot 8^2))$ as functions of strip width $L_y$.}}
\begin{center}
\begin{tabular}{||l|l|l|l||}
\hline
$(L_y+1)/L_y$ & $[\lambda_{(4 \cdot 8^2),L_y+1,free}(-1)/
\lambda_{(4 \cdot8^2),L_y,free}(-1)]^{1/4}$ & 
$R_{(4 \cdot 8^2),\frac{L_y^2}{(L_y-1)(L_y+1)},free}(-1)$ \\ \hline\hline
2/1 & $\frac{(889)^{1/4}}{2}=2.73020617$  & \\ \hline
3/2 & 2.73010279  & 1.00003787 \\ \hline
4/3 & 2.73009404  & 1.00000320 \\ \hline
5/4 & 2.73009323  & 1.00000030 \\ \hline
6/5 & 2.730093140 & 1.000000032 \\ \hline
\end{tabular}
\end{center}
\label{upperbounds_alpha_488_table}
\end{table}


\begin{table}
  \caption{\footnotesize{Lower bounds and their ratios for 
$\alpha_0((4 \cdot 8^2))$ as functions of strip width $L_y$.}}
\begin{center}
\begin{tabular}{||l|l|l|l||}
\hline
BC  & $L_y$ & $[\lambda_{(4 \cdot 8^2),L_y,free/cyl}(0)]^{1/(4L_y)}$ &
$R_{(4 \cdot 8^2),\frac{L_y+1}{L_y}/\frac{L_y+2}{L_y},free/cyl}(0)$ \\ 
\hline\hline
free  & 2  & $(21)^{1/8}=1.46311146$  & \\ \hline
free  & 3  & 1.65063068            & 1.12816469 \\ \hline
free  & 4  & 1.75020633            & 1.06032582 \\ \hline
free  & 5  & 1.81176342            & 1.03517133 \\ \hline
free  & 6  & 1.85353652            & 1.02305659 \\ \hline\hline
cyl   & 2  & $\sqrt{3}=1.73205081$ & \\ \hline
cyl   & 4  & 1.98451595            & 1.14576082 \\ \hline
cyl   & 6  & 2.032649948           & 1.024254778 \\ \hline
\end{tabular}
\end{center}
\label{lowerbounds_alpha0_488_table}
\end{table}


\begin{table}
  \caption{\footnotesize{Upper bounds and their ratios for 
$\alpha_0((4 \cdot 8^2))$ as functions of strip width $L_y$.}}
\begin{center}
\begin{tabular}{||l|l|l|l||}
\hline
$(L_y+1)/L_y$ & $[\lambda_{(4 \cdot 8^2),L_y+1,free}(0)/
\lambda_{(4 \cdot 8^2), L_y,free}(0)]^{1/4}$ & 
$R_{(4 \cdot 8^2),\frac{L_y^2}{(L_y-1)(L_y+1)},free}(0)$ \\ \hline\hline
2/1 & $(21)^{1/4}=2.14069514$             & \\ \hline
3/2 & 2.10084938  & 1.01896650 \\ \hline
4/3 & 2.08644655  & 1.00690304 \\ \hline
5/4 & 2.08041720  & 1.00289815 \\ \hline
6/5 & 2.077301063 & 1.001500087 \\ \hline
\end{tabular}
\end{center}
\label{upperbounds_alpha0_488_table}
\end{table}


\begin{table}
  \caption{\footnotesize{Lower bounds and their ratios for 
$\beta((4 \cdot 8^2))$ as 
functions of strip width $L_y$. The abbreviation tor stands for
``toroidal''.}} 
\begin{center}
\begin{tabular}{||l|l|l|l||}
\hline
BC & $L_y$ &
$[\lambda_{(4 \cdot8^2) L_y,cyc/tor}(-1,1)]^{1/(4L_y)}$ &
$R_{(4 \cdot 8^2), \frac{L_y+1}{L_y}/\frac{L_y+2}{L_y}, cyc/tor}(-1,1)$ \\ 
\hline\hline
cyc & 2  & $\sqrt{2} = 1.41421356$ & \\ \hline
cyc & 3  & 1.62047257   & 1.14584715 \\ \hline
cyc & 4  & 1.73110235   & 1.06827007 \\ \hline
cyc & 5  & 1.80061384   & 1.04015446 \\ \hline\hline
tor & 2  & 2            &  \\ \hline
tor & 4  & 2.080338691  & 1.040169345 \\ \hline
\end{tabular}
\end{center}
\label{lowerbounds_beta_488_table}
\end{table}


\begin{table}
  \caption{\footnotesize{Upper bounds and their ratios for 
$\beta((4 \cdot 8^2))$ as functions of strip width $L_y$.}} 
\begin{center}
\begin{tabular}{||l|l|l|l||}
\hline
$(L_y+1)/L_y$ & $\big [ \frac{\lambda_{(4 \cdot 8^2), L_y+1, cyc} (-1,1)}
{\lambda_{(4 \cdot 8^2), L_y, cyc} (-1,1)} \big ]^{1/4}$ & 
$R_{(4 \cdot 8^2), \frac{L_y^2}{(L_y-1)(L_y+1)}, cyc} (-1,1)$ \\ \hline\hline
3/2 & 2.12762488  &  \\ \hline
4/3 & 2.11040559  & 1.00815923  \\ \hline
5/4 & 2.107715225 & 1.001276438 \\ \hline
\end{tabular}
\end{center}
\label{upperbounds_beta_488_table}
\end{table}


\begin{table}
  \caption{\footnotesize{Lower bounds and their ratios for 
$\alpha(kag)$ as functions of strip width $L_y$. In this table and the others,
the abbreviation cyl stands for ``cylindrical''.}} 
\begin{center}
\begin{tabular}{||l|l|l|l||}
\hline
BC   & $L_y$ & $[\lambda_{kag, L_y, free/cyl}(-1)]^{1/(3L_y)}$ & 
$R_{kag, (L_y+1)/L_y, free/cyl} (-1)$ \\ \hline\hline
free   & 1  & $(12)^{1/3} = 2.289428485$ & \\ \hline
free   & 2  & $3^{1/3}\Big( \frac{47+\sqrt{2113}}{2} \Big )^{1/6}=2.73478917$
                                  & 1.172619975 \\ \hline
free   & 3  & 2.9014136165        & 1.07624544 \\ \hline\hline
cyl    & 1  & $(18)^{1/3}=2.62074139$ & \\ \hline
cyl    &2& $[6(88+\sqrt{7609} \, )]^{1/6}=3.18879387$  & 1.216752585 \\ \hline
cyl    & 3  & 3.2490590695 & 1.018899059 \\ \hline
\end{tabular}
\end{center}
\label{lowerbounds_alpha_kag_table}
\end{table}


\begin{table}
  \caption{\footnotesize{Upper bounds and their ratios for 
$\alpha(kag)$ as functions of strip width $L_y$.}} 
\begin{center}
\begin{tabular}{||l|l|l||}
\hline
$(L_y+1)/L_y$&$(\lambda_{kag,L_y+1,free}(-1)/\lambda_{kag,L_y,free}(-1))^{1/3}$
& $R_{kag,\frac{L_y^2}{(L_y-1)(L_y+1)},free} (-1)$ \\ \hline\hline
2/1 & $\Big [\frac{3(47+\sqrt{2113}\, )}{8}\Big ]^{1/3} = 3.26678550$ & \\ 
\hline
3/2 & 3.2657371991  & 1.000321000 \\ \hline
\end{tabular}
\end{center}
\label{upperbounds_alpha_kag_table}
\end{table}


\begin{table}
  \caption{\footnotesize{Lower bounds on $\alpha_0(kag)$ and their ratios,
      as functions of strip width $L_y$.}}
\begin{center}
\begin{tabular}{||l|l|l|l||}
\hline
BC  & $L_y$ & $[\lambda_{kag,L_y,free/cyl}(0)]^{1/(3L_y)}$ &
$R_{kag, (L_y+1)/L_y, free/cyl}(0)$ \\ \hline\hline
free & 1   & $2^{1/3} = 1.25992105$ &  \\ \hline
free & 2   & $2^{1/3}(5+\sqrt{23} \, )^{1/6}=1.84296413$ & 1.43903358 \\ \hline
free & 3   & 2.07555502            & 1.15009316 \\ \hline\hline
cyl  & 1   & $2^{2/3}=1.58740105.$ & \\ \hline
cyl  & 2   & $[2(3+\sqrt{11} \, )]^{1/3}=2.32901182$ & 1.46718551 \\ \hline
cyl  & 3   & 2.481974714 & 1.065677167 \\ \hline
\end{tabular}
\end{center}
\label{lowerbounds_alpha0_kag_table}
\end{table}
         

\begin{table}
  \caption{\footnotesize{Upper bounds on $\alpha_0(kag)$ and their ratios, 
      as functions of strip width $L_y$.}} 
\begin{center}
\begin{tabular}{||l|l|l||}
\hline
$(L_y+1)/L_y$&$[\lambda_{kag,L_y+1,free}(0)/\lambda_{kag,L_y,free}(0)]^{1/3}$ 
& $R_{kag,\frac{L_y^2}{(L_y-1)(L_y+1)}, free}(0)$           \\ \hline\hline
2/1 & $[2(5+\sqrt{23})]^{1/3}=2.695817165$  &                   \\ \hline
3/2 & 2.632503652                           & 1.024050684       \\ \hline
\end{tabular}
\end{center}
\label{upperbounds_alpha0_kag_table}
\end{table}


\begin{table}
  \caption{\footnotesize{Lower bounds and their ratios for 
$\beta(kag)$ as functions of strip width $L_y$.}} 
\begin{center}
\begin{tabular}{||l|l|l|l||}
\hline
BC  & $L_y$ & $[\lambda_{kag, L_y, cyc}(-1,1)]^{1/(3L_y-1)}$ or
$[\lambda_{kag, L_y, tor}(-1,1)]^{1/(3L_y)}$ & 
$R_{kag, \frac{L_y+1}{L_y},cyc/tor} (-1,1)$ \\ \hline\hline
cyc & 2  & $(55)^{1/5} = 2.22880738$  & \\ \hline
cyc & 3  & 2.653725025                & 1.19064799 \\ \hline\hline
tor & 1  & $(30)^{1/3}=3.10723251$    & \\ \hline
tor & 2  & $[10((79 + 2\sqrt{1585})])^{1/6}=3.415032724$ & 1.099059281 \\ 
\hline
\end{tabular}
\end{center}
\label{lowerbounds_beta_kag_table}
\end{table}


\begin{table}
  \caption{\footnotesize{Upper bounds and their ratios for 
$\beta(kag)$ as functions of strip width $L_y$.}} 
\begin{center}
\begin{tabular}{||l|l|l|l||}
\hline
$(L_y+1)/L_y$ & $\big [\frac{\lambda_{kag, L_y+1, cyc} (-1,1)} 
{\lambda_{kag,L_y, cyc} (-1,1)} \big ]^{1/3}$ & 
$R_{kag,\frac{L_y^2}{(L_y-1)(L_y+1)}, cyc} (-1,1)$ \\ \hline\hline
2/1 & $(55)^{1/3} = 3.80295246$  & \\ \hline
3/2 & 3.549454037                & 1.071418990. \\ \hline
\end{tabular}
\end{center}
\label{upperbounds_beta_kag_table}
\end{table}


\begin{table}
  \caption{\footnotesize{Lower bounds on $\alpha((3^3 \cdot 4^2))$ and 
  their ratios, as functions of strip width $L_y$.}} 
\begin{center}
\begin{tabular}{||l|l|l|l||}
\hline
BC & $L_y$ & $[\lambda_{(3^3 \cdot 4^2),L_y,free/cyl}(-1)]^{1/(2L_y)}$ &
$R_{(3^3.4^2),\frac{L_y+2}{L_y} /\frac{L_y+1}{L_y},free/cyl}(-1)$ \\ 
\hline\hline
free & 3 & $(17+4\sqrt{13})^{1/3}$      & \\
     &   & $ = 3.15557776$              & \\ \hline
free & 5 & 3.45440528                   & 1.09469820 \\ \hline\hline
cyl  & 2 & 3                            & \\ \hline
cyl  & 4 & 3.82776685                   & 1.27592228  \\ \hline
cyl  & 6 & 3.922582062                  & 1.024770372  \\ \hline\hline
free & 2 & $\sqrt{3} \times 7^{1/4}=2.81731325$  & \\
\hline
free & 3 & $(497 + \sqrt{240313})^{1/6}$ & 1.12004032 \\
     &   & $ = 3.15550442$               & \\ \hline
free & 4 & 3.33914866                    & 1.05819806 \\ \hline
free & 5 & 3.45434518                    & 1.03449877 \\ \hline
free & 6 & 3.53332068                    & 1.02286265 \\ \hline\hline
cyl  & 2 & $(4 \times 21)^{1/4}=3.0274001$  & \\
\hline
cyl  & 3 & $(2414)^{1/6}=3.66260045$     & 1.2098171 \\ \hline
cyl  & 4 & $[6(3909 + 13\sqrt{89841})]^{1/8}$&1.04720046 \\
     &   & $ = 3.83547688$                 & \\ \hline
cyl  & 5 & $(407837 + 5 \sqrt{6475806457})^{1/10}$ & 1.016341829 \\
     &   & $ = 3.898155587$                & \\ \hline
\end{tabular}
\end{center}
\label{lowerbounds_alpha_33344_table}
\end{table}


\begin{table}
  \caption{\footnotesize{Upper bounds on $\alpha((3^3 \cdot 4^2))$ and 
  their ratios, as functions of strip width $L_y$.}} 
\begin{center}
\begin{tabular}{||l|l|l|l||}
\hline
$\frac{L_y+2}{L_y}$ or $\frac{L_y+1}{L_y}$ & 
$\sqrt{ \frac{\lambda_{(3^3 \cdot 4^2),L_y+2/1,free}(-1)}
             {\lambda_{(3^3 \cdot 4^2),L_y,free}(-1)} }$ & 
$R_{(3^3 \cdot 4^2),
\frac{L_y^2}{(L_y-2)(L_y+2)}/\frac{L_y^2}{(L_y-1)(L_y+1)},free}(-1)$ \\ 
\hline\hline
3/1 & $\frac{2+\sqrt{13}}{\sqrt{2}}=3.96372332$  & \\ \hline
5/3 & 3.95653392 & 1.001817095 \\ \hline\hline
2/1 & $\frac{\sqrt{63}}{2}=3.96862697$  & \\ \hline
3/2 & $\frac{1}{3}\sqrt{71+\frac{\sqrt{240313}}{7}}$ & 1.00254569 \\
    & $ = 3.9585497$                                 & \\ \hline
4/3 & 3.95673204  & 1.00045939 \\ \hline
5/4 & 3.95626750  & 1.00011742 \\ \hline
6/5 & 3.956121920 & 1.000036798 \\ \hline
\end{tabular}
\end{center}
\label{upperbounds_alpha_33344_table}
\end{table}


\begin{table}
  \caption{\footnotesize{Lower bounds on $\alpha_0((3^3 \cdot 4^2))$ and 
  their ratios, as functions of strip width $L_y$.}} 
\begin{center}
\begin{tabular}{||l|l|l|l||}
\hline
BC & $L_y$ & $[\lambda_{(3^3 \cdot 4^2),L_y,free/cyl}(0)]^{1/(2L_y)}$ &
$R_{(3^3.4^2), \frac{L_y+2}{L_y} /\frac{L_y+1}{L_y},free/cyl}(0)$ \\ 
\hline\hline
free & 3  & $\big( \frac{13+\sqrt{105}}{2} \big)^{1/3}=2.2652284$ & \\ \hline
free & 5  & 2.63777102        & 1.16446139 \\ \hline\hline
cyl  & 2  & 2  & \\ \hline
cyl  & 4  & 2.95888008        & 1.47944004 \\ \hline
cyl  & 6  & 3.142411228       & 1.062027234 \\ \hline\hline
free & 2  & $(4 \times 3)^{1/4}=1.86120972$  & \\
\hline
free & 3  & $\big( \frac{137+\sqrt{17713}}{2} \big)^{1/6}=2.26506049$ 
                              & 1.21698295 \\ \hline
free & 4     & 2.49249353     & 1.10040926 \\ \hline
free & 5     & 2.63744871     & 1.05815669 \\ \hline
free & 6     & 2.73767800     & 1.03800237 \\ \hline\hline
cyl  & 2     & $(9 \times 2)^{1/4}=2.05976714$  & \\
\hline
cyl  & 3     & $2^{5/6} \times 13^{1/6}=2.73221930$ & 1.32646999 \\ \hline
cyl  & 4     & $(3096 + 6\sqrt{264981})^{1/8}$      & 1.08992900 \\
     &       & $ = 2.97792504$                      & \\ \hline
cyl  & 5     & $(39973 + \sqrt{1566836161})^{1/10}$ & 1.037895636 \\
     &       & $=3.09077540$  & \\ \hline
\end{tabular}
\end{center}
\label{lowerbounds_alpha0_33344_table}
\end{table}


\begin{table}
  \caption{\footnotesize{Upper bounds on $\alpha_0((3^3 \cdot 4^2))$ and 
  their ratios, as functions of strip width $L_y$.}} 
\begin{center}
\begin{tabular}{||l|l|l|l||}
\hline
$\frac{L_y+2}{L_y}$ or $\frac{L_y+1}{L_y}$ & 
$\sqrt{ \frac{\lambda_{(3^3 \cdot 4^2),L_y+2/1,free}(0)}
             {\lambda_{(3^3 \cdot 4^2),L_y,free}(0)}}$ &
$R_{(3^3 \cdot 4^2), \frac{L_y^2}{(L_y-2)(L_y+2)}/
\frac{L_y^2}{(L_y-1)(L_y+1)},free} (0)$ \\ \hline\hline
3/1 & $\frac{\sqrt{21}+\sqrt{5}}{2}=3.40932184$  & \\ \hline
5/3 & 3.31455119 & 1.0285923 \\ \hline\hline
2/1 & $2\sqrt{3}=3.464101615$  & \\ \hline
3/2 & $\frac{1}{2} \sqrt{\frac{137+\sqrt{17713}}{6}}$ & 1.03262310 \\
    & $ = 3.35466215$  & \\ \hline
4/3 & 3.32121311  & 1.01007133 \\ \hline
5/4 & 3.30661746  & 1.00441407 \\ \hline
6/5 & 3.298937504 & 1.002328009 \\ \hline
\end{tabular}
\end{center}
\label{upperbounds_alpha0_33344_table}
\end{table}


\begin{table}
  \caption{\footnotesize{Lower bounds and their ratios for 
$\beta(3^3 \cdot 4^2)$ as functions of strip width $L_y$.}} 
\begin{center}
\begin{tabular}{||l|l|l|l||}
\hline
BC  & $L_y$ & $\lambda_{(3^3 \cdot 4^2),L_y,cyc/tor}^{1/(2L_y)}(-1,1)$ &
$R_{(3^3 \cdot 4^2),\frac{L_y+2}{L_y}/ \frac{L_y+1}{L_y}, cyc/tor} (-1,1)$ \\ 
\hline\hline
cyc  & 3  & 3.32494691                       & \\ \hline
cyc  & 5  & 4.03289325                       & 1.21291959 \\ \hline\hline
tor  & 2  & $\frac{\sqrt{2(25+\sqrt{613})}}{2}=4.987927265$ & \\
\hline
tor  & 4  & 5.26288016                       & 1.05512368 \\ \hline\hline
cyc  & 2  & $(43)^{1/4} = 2.5607496$         &            \\ \hline
cyc  & 3  & 3.31487994                       & 1.2944959 \\ \hline
cyc  & 4  & 3.74829168                       & 1.1307473 \\ \hline
cyc  & 5  & 4.02663152                       & 1.0742578 \\ \hline\hline
tor  & 2  & $(584)^{1/4}=4.91590195$         &           \\ \hline
tor  & 3  & 5.176853205                      & 1.05308309 \\ \hline
\end{tabular}
\end{center}
\label{lowerbounds_beta_33344_table}
\end{table}


\clearpage

\begin{table}
  \caption{\footnotesize{Upper bounds and their ratios for 
$\beta(3^3 \cdot 4^2)$ as functions of strip width $L_y$.}} 
\begin{center}
\begin{tabular}{||l|l|l|l||}
\hline
$\frac{L_y+2}{L_y}$ or $\frac{L_y+1}{L_y}$ & 
$\sqrt{ \frac{\lambda_{(3^3 \cdot 4^2), L_y+2/1, cyc} (-1,1)} 
      {\lambda_{ (3^3 \cdot 4^2), L_y, cyc} (-1,1)} }$ &
$R_{(3^3 \cdot 4^2),\frac{L_y^2}{(L_y-2)(L_y+2)}/
\frac{L_y^2}{(L_y-1)(L_y+1)},cyc}(-1,1)$ \\ 
\hline\hline
3/1  & 6.06285349                         & \\ \hline
5/3  & 5.38722039                         & 1.12541404 \\ \hline\hline
2/1  & $\sqrt{43} = 6.5574385$            &            \\ \hline
3/2  & 5.55480969                         & 1.18049742 \\ \hline
4/3  & 5.41913669                         & 1.02503591 \\ \hline
5/4  & 5.362606470                        & 1.010541556 \\ \hline
\end{tabular}
\end{center}
\label{upperbounds_beta_33344_table}
\end{table}


\begin{table}
  \caption{\footnotesize{Lower bounds on 
$\alpha((3^2 \cdot 4 \cdot 3 \cdot 4))$ and 
  their ratios, as functions of strip width $L_y$.}} 
\begin{center}
\begin{tabular}{||l|l|l|l||}
\hline
BC & $L_y$ & 
$[\lambda_{(3^2 \cdot 4 \cdot 3 \cdot 4),L_y,free/cyl}(-1)]^{1/(2L_y)}$ &
$R_{(3^2 \cdot 4 \cdot 3 \cdot 4),\frac{L_y+1}{L_y}/\frac{L_y+2}{L_y},
free/cyl}(-1)$ \\ \hline\hline
free  & 2  & $(9 \times 7)^{1/4}=2.81731325$ & \\
\hline
free  & 3  & $(17+4\sqrt{13})^{1/3}$ & 1.12006635 \\
      &    & $ = 3.15557776$ & \\ \hline
free  & 4  & 3.33926081  & 1.05820901 \\ \hline
free  & 5  & 3.45448103  & 1.03450471 \\ \hline
free  & 6  & 3.53347262  & 1.02286641 \\ \hline\hline
cyl   & 2  & 3           &            \\ \hline
cyl   & 4  & 3.82776685  & 1.27592228 \\ \hline
cyl   & 6  & 3.922582062 & 1.024770372 \\ \hline
\end{tabular}
\end{center}
\label{lowerbounds_alpha_33434_table}
\end{table}


\begin{table}
  \caption{\footnotesize{Upper bounds on 
$\alpha((3^2 \cdot 4 \cdot 3 \cdot 4))$ and 
  their ratios, as functions of strip width $L_y$.}} 
\begin{center}
\begin{tabular}{||l|l|l||}
\hline
$(L_y+1)/L_y$ & 
$\sqrt{\frac{\lambda_{(3^2 \cdot 4 \cdot 3 \cdot 4),L_y+1,free}(-1)}
{\lambda_{(3^2 \cdot 4 \cdot 3 \cdot 4),L_y,free}(-1)}}$ & 
$R_{(3^2 \cdot 4 \cdot 3 \cdot 4),\frac{L_y^2}{(L_y-1)(L_y+1)},free}(-1)$ \\ 
\hline\hline
2/1 & $\frac{3\sqrt{7}}{2}=3.96862697$ & \\ \hline
3/2 & $\frac{17+4\sqrt{13}}{3\sqrt{7}}=3.9588257$ & 1.00247579 \\ \hline
4/3 & 3.95698775  & 1.00046449 \\ \hline
5/4 & 3.95651388  & 1.00011977 \\ \hline
6/5 & 3.956364741 & 1.000037697 \\ \hline
\end{tabular}
\end{center}
\label{upperbounds_alpha_33434_table}
\end{table}


\begin{table}
  \caption{\footnotesize{Lower bounds on 
$\alpha_0((3^2 \cdot 4 \cdot 3 \cdot 4))$ and 
  their ratios, as functions of strip width $L_y$.}} 
\begin{center}
\begin{tabular}{||l|l|l|l||}
\hline
BC  & $L_y$ & 
$[\lambda_{(3^2 \cdot 4 \cdot 3 \cdot 4),L_y,free/cyl}(0)]^{1/(2L_y)}$ &
$R_{(3^2\cdot 4\cdot 3\cdot 4),\frac{L_y+1}{L_y}/\frac{L_y+2}{L_y},free/cyl}(0)$ \\ \hline\hline
free  & 2  & $(4 \times 3)^{1/4}=1.86120972$ & \\
\hline
free  & 3  & $\big( \frac{13+\sqrt{105}}{2} \big)^{1/3}=2.26522841$ 
                         & 1.21707317  \\ \hline
free  & 4  & 2.49270953  & 1.10042304  \\ \hline
free  & 5  & 2.63768181  & 1.05815851  \\ \hline
free  & 6  & 2.73791775  & 1.03800153  \\ \hline\hline
cyl   & 2  & 2           &             \\ \hline
cyl   & 4  & 2.95888008  & 1.47944004  \\ \hline
cyl   & 6  & 3.142411229 & 1.062027234  \\ \hline
\end{tabular}
\end{center}
\label{lowerbounds_alpha0_33434_table}
\end{table}


\begin{table}
  \caption{\footnotesize{Upper bounds on 
$\alpha_0(n(3^2 \cdot 4 \cdot 3 \cdot 4))$ and 
  their ratios, as functions of strip width $L_y$.}} 
\begin{center}
\begin{tabular}{||l|l|l||}
\hline
$(L_y+1)/L_y$ & 
$\sqrt{ \frac{\lambda_{(3^2 \cdot 4 \cdot 3 \cdot 4),L_y+1,free}(0)}
{\lambda_{(3^2 \cdot 4 \cdot 3 \cdot 4),L_y,free}(0)} }$ & 
$R_{(3^2 \cdot 4 \cdot 3 \cdot 4),\frac{L_y^2}{(L_y-1)(L_y+1)},free} (0)$ \\ 
\hline\hline
2/1 & $2\sqrt{3}=3.464101615$  & \\ \hline
3/2 & $\frac{13+\sqrt{105}}{4\sqrt{3}}=3.35540832$ & 1.03239346 \\ \hline
4/3 & 3.32162574  & 1.01017050 \\ \hline
5/4 & 3.30693243  & 1.00444318 \\ \hline
6/5 & 3.299213098 & 1.002339750 \\ \hline
\end{tabular}
\end{center}
\label{upperbounds_alpha0_33434_table}
\end{table}


\begin{table}
  \caption{\footnotesize{Lower bounds and their ratios for 
$\beta(3^2 \cdot 4 \cdot 3 \cdot 4)$ as functions of strip width $L_y$.}} 
\begin{center}
\begin{tabular}{||l|l|l|l||}
\hline
BC  & $L_y$ & 
$[\lambda_{(3^2 \cdot 4 \cdot 3 \cdot 4),L_y,cyc/tor}(-1,1)]^{1/(2L_y)}$ &
$R_{(3^2 \cdot 4 \cdot 3 \cdot 4),\frac{L_y+1}{L_y}/\frac{L_y+2}{L_y},cyc/tor}
(-1,1)$ \\ \hline\hline
cyc  & 2  & $(43)^{1/4}=2.56074960$      & \\ \hline
cyc  & 3  & 3.32202041                    & 1.29728436 \\ \hline
cyc  & 4  & 3.75433393                    & 1.13013572 \\ \hline
cyc  & 5  & 4.03143679                    & 1.07380879 \\ \hline\hline
tor  & 2  & $\frac{\sqrt{2(25+\sqrt{613})}}{2}=4.987927265$ & \\
 \hline
tor  & 4  & 5.264056522                   & 1.055359520 \\ \hline
\end{tabular}
\end{center}
\label{lowerbounds_beta_33434_table}
\end{table}


\begin{table}
  \caption{\footnotesize{Upper bounds and their ratios for 
$\beta(3^2 \cdot 4 \cdot 3 \cdot 4)$ as functions of strip width $L_y$.}} 
\begin{center}
\begin{tabular}{||l|l|l|l||}
\hline
$\frac{L_y+1}{L_y}$ & 
$\sqrt{\frac{\lambda_{(3^2 \cdot 4 \cdot 3 \cdot 4),L_y+1,cyc} (-1,1)}
            {\lambda_{(3^2 \cdot 4 \cdot 3 \cdot 4),L_y,cyc}(-1,1)}}$ & 
$R_{(3^2 \cdot 4 \cdot 3 \cdot 4),\frac{L_y^2}{(L_y-1)(L_y+1)},cyc}(-1,1)$ \\ 
\hline\hline
2/1 & $\sqrt{43} = 6.55743852$     &            \\ \hline
3/2 & 5.59078335                   & 1.17290156 \\ \hline
4/3 & 5.41906930                   & 1.03168700 \\ \hline
5/4 & 5.360035653                  & 1.011013668 \\ \hline
\end{tabular}
\end{center}
\label{upperbounds_beta_33434_table}
\end{table}


\begin{table}
\caption{\footnotesize{Values of the exponential growth constants (EGCs)
    $\alpha(\Lambda)$, $\alpha_0(\Lambda)$, and 
    $\beta(\Lambda)$ for the Archimedean lattices
    $\Lambda$ analyzed here via calculations on sequences of finite-width, 
    infinite-length strips. In the right-most column we list the exactly
    known values of $\tau(\Lambda)$.  The lattices are listed
    in order of increasing vertex degree $\Delta(\Lambda)$ and, for a given 
    vertex degree, in order of increasing girth, $g(\Lambda)$. 
    For the EGCs that are not exactly known, we list the approximate values
    that we have obtained from our upper and lower bounds, as defined in
    Eq. (\ref{xi_value}). In the case of the homopolygonal lattices, 
    (hc), (sq), and (tri), we list the exact values of ($\alpha(tri)$, 
    $\alpha_0(tri)$, and $\beta(hc)$ and the approximate values of the other
    EGCs that we obtained in \cite{ac}. See text for further discussion.}}
\begin{center}
\begin{tabular}{|c|c|c|c|c|c|c|} \hline\hline
$\Lambda$ & $\Delta(\Lambda)$ & $g(\Lambda)$ & $\alpha(\Lambda)$ & 
$\alpha_0(\Lambda)$ & $\beta(\Lambda)$ & $\tau(\Lambda)$\\
\hline
$(4 \cdot 8^2)$  & 3 & 4       & $2.7299 \pm 0.0002$   & $2.055 \pm 0.022$  
                               & $2.094 \pm 0.014$  & 2.196103   \\
$(6^3)=$ hc      & 3 & 6       & $2.78284 \pm 0.00064$ & $2.134 \pm 0.027$ 
                               & 2.115336           & 2.242665    \\
\hline
$(3\cdot 6\cdot 3\cdot 6)$&4&3 & $3.2574 \pm 0.0083$   & $2.557\pm 0.075$  
                               & $3.482 \pm 0.067$  & 3.113341    \\
$(4^4)=$ sq      & 4 & 4       & $3.49359 \pm 0.00034$ & $2.846 \pm 0.016$  
                               & $3.49359 \pm 0.00034$ & 3.209912 \\ 
\hline
$(3^3 \cdot 4^2)$& 5 & 3       & $3.939 \pm 0.017$     & $3.221 \pm 0.078$
                               & $5.313 \pm 0.050$     & 4.083383  \\ 
$(3^2 \cdot 4 \cdot 3 \cdot 4)$
                 & 5 & 3       & $3.939 \pm 0.017$     & $3.221 \pm 0.078$   
                               & $5.312 \pm 0.048$     & 4.099462   \\
\hline 
$(3^6)=$ tri     & 6 & 3       & 4.474647              & 3.770920
                               & $7.7442 \pm 0.0036$   & 5.029546    \\
\hline\hline
\end{tabular}
\end{center}
\label{egc_values_table}
\end{table}


\begin{table}
  \caption{\footnotesize{Values of $\alpha_{u,w}(\Lambda)$, 
$\alpha_{0,u,w}(\Lambda)$, and $\beta_{u,w}(\Lambda)$ or
      $\beta_{u,w'}(\Lambda)$ for Archimedean lattices $\Lambda$. 
  The last column lists the (exactly known)
  values of $\tau(\Lambda)$. See text for definitions and notation.}} 
\begin{center}
\begin{tabular}{|c|c|c|c|c|c|c|} \hline\hline
$\Lambda$ & $\Delta(\Lambda)$ & $g(\Lambda)$ & $\alpha_{u,w}(\Lambda)$ &  
$\alpha_{0,u,w}(\Lambda)$ & $\beta_{u,w}(\Lambda)$ & $\tau(\Lambda)$  \\ 
\hline
$(3 \cdot 12^2)$  & 3 & 3 & 2.569587 & 1.878922 & 2.039649 & 2.055591 \\
$(4 \cdot 8^2)$   & 3 & 4 & 2.730206 & 2.140695 & 2.101400 & 2.196103  \\
$(4\cdot 6\cdot 12)$&3& 4 & 2.721014 & 2.097345 & 2.108019 & 2.1766685  \\

$(6^3)=$ hc      & 3 & 6 & 2.783882 & 2.236068  & 2.121320 & 2.242665  \\
\hline 
$(3 \cdot 6 \cdot 3 \cdot 6)$
                  & 4 & 3 & 3.267168 & 2.714418 & 3.5      & 3.113341 \\
$(3 \cdot 4 \cdot 6 \cdot 4)$ 
                  & 4 & 3 & 3.381580 & 2.853639 & 3.5      & 3.141816  \\
$(4^4)=$ sq       & 4 & 4 & 3.5      & 3        & 3.5      & 3.209912  \\
\hline
$(3^3 \cdot 4^2)$ & 5 & 3 & 3.968627 & 3.464102 & 5.303301 & 4.083383  \\
$(3^2 \cdot 4 \cdot 3 \cdot 4)$ 
                  & 5 & 3 & 3.968627 & 3.464102 & 5.303301 & 4.099462  \\
$(3^4 \cdot 6)$   & 5 & 3 & 3.834352 & 3.295098 & 5.303301 & 4.022983  \\
\hline
$(3)^6=$ tri     & 6 & 3 & 4.5      & 4        & 7.75      & 5.029546  \\
\hline\hline
\end{tabular}
\end{center}
\label{uw_table}
\end{table}


\begin{table}
  \caption{\footnotesize{Values of $\alpha_{u,w}(\Lambda_{dual})$, 
   $\alpha_{0,u,w}(\Lambda_{dual})$, and $\beta_{u,w}(\Lambda_{dual})$ for
   duals of Archimedean lattices, $\Lambda_{dual}$. 
   For the $[3 \cdot 12^2]$, $[4 \cdot 8^2]$, and $[4 \cdot 6 \cdot 12]$ 
   lattices we list the values of $\alpha_{u,w'}$ and $\alpha_{0,u,w'}$. 
   In the last column we list the (exactly known) values of 
   $\tau(\Lambda_{dual})$. See text for definitions and notation.}} 
\begin{center}
\begin{tabular}{|c|c|c|c|c|c|c|} \hline\hline
$\Lambda_{dual}$ & $p(\Lambda_{dual})$ & $\Delta_{eff}(\Lambda_{dual})$ & 
$\alpha_{u,w}(\Lambda_{dual})$ &  $\alpha_{0,u,w}(\Lambda_{dual})$ & 
$\beta_{u,w}(\Lambda_{dual})$ & $\tau(\Lambda_{dual})$ \\ 
\hline
$[3 \cdot 12^2]$  & 3 & 6 & 4.160168 & 3.301927 & 6.6027795& 4.225454  \\
$[4 \cdot 8^2]$   & 3 & 6 & 4.415880 & 3.741657 & 7.454026 & 4.822867  \\
$[4\cdot 6\cdot 12]$&3& 6 & 4.443744 & 3.825862 & 7.403934 & 4.737886  \\

$[6^3]=$ tri      & 3 & 6 & 4.5      & 4         & 7.75    & 5.029546  \\
\hline 
$[3 \cdot 6 \cdot 3 \cdot 6]$
                  & 4 & 4 & 3.5   & 3  & 3.2671675  & 3.113341   \\
$[3 \cdot 4 \cdot 6 \cdot 4]$ 
                  & 4 & 4 & 3.5   & 3  & 3.381580   & 3.141816   \\
$[4^4]=$ sq       & 4 & 4 & 3.5   & 3  & 3.5        & 3.209912   \\
\hline
$[3^3 \cdot 4^2]$ & 5 &10/3 & 3.041101 & 2.519842 & 2.506649 & 2.554740 \\
$[3^2 \cdot 4 \cdot 3 \cdot 4]$ 
                  & 5 &10/3 & 3.041101 & 2.519842 & 2.506649 & 2.561442 \\
$[3^4 \cdot 6]$   & 5 &10/3 & 3.041101 & 2.519842 & 2.449785 & 2.529485 \\
\hline
$[3^6]=$ hc       & 6 & 3   & 2.783882 & 2.236068 & 2.121320 & 2.242665 \\
\hline\hline
\end{tabular}
\end{center}
\label{uw_duals_table}
\end{table}


\end{document}